\documentclass[preprint] {article}
\usepackage{amsmath,amssymb,amsfonts}
\usepackage{geometry}
\usepackage{graphicx}
\usepackage{subcaption}
\usepackage{braket}
\usepackage{float}
\allowdisplaybreaks
\newcommand {\bp}{\begin{pmatrix}}
\newcommand {\ep}{\end{pmatrix}}
\newcommand{\be}{\begin{equation}} \newcommand{\ee}{\end{equation}}
\newcommand{\bea}{\begin{eqnarray}}\newcommand{\eea}{\end{eqnarray}}

\begin{document}
\title{ Balanced loss-gain induced chaos in a periodic Toda lattice}

\author{ Puspendu Roy\footnote {{\bf email}:puspenduroy716@gmail.com}
\ and \ Pijush K. Ghosh\footnote{{\bf email:} 
pijushkanti.ghosh@visva-bharati.ac.in}
\footnote{Corresponding Author} 
}
\date{Department of Physics, Siksha-Bhavana, \\ 
Visva-Bharati University, \\
Santiniketan, PIN 731 235, India.}

\maketitle
\begin{abstract}

We consider equal-mass periodic Toda oscillators with balanced loss-gain for
two and three particles. The two-particle system is integrable with the
Hamiltonian and the genralized total momentum being two integrals of motion.
The model in its full generality is not amenable to exact analytic solutions,
and investigated numerically showing existence of regular periodic solutions.
The three-particle equal-mass periodic Toda lattice is considered in presence
of balanced loss-gain and velocity mediated coupling.
The system is Hamiltonian for the special case whenever the strength of the
velocity-mediated coupling is half of the strength of the loss-gain. The model
admits regular and chaotic solutions for vanishing as well as non-vanishing
velocity-mediated coupling, including the Hamiltonian system. The chaos is
induced due to the presence of balanced loss-gain, since undriven
equal-mass Toda lattice is non-chaotic.
The chaotic behaviour is studied in detail for the Hamiltonian
system as well as for the system with vanishing velocity mediated coupling
by using time series, Poincar\'e sections, auto-correlation function,
power spectra, Lyapunov exponent and bifurcation diagram.
	
\end{abstract}
\section{Introduction}

The Toda lattice is one of the most important integrable systems with
various applications in physical systems\cite{mt,toda-book}. The integrability\cite{flascha,mh,mpc},
the existence of periodic and stable solitary wave solutions\cite{kac,tanaka}, exact solutions for the
dynamics, and statistical thermodynamics\cite{ac} are some of the most crucial
characteristics of a Toda lattice. The applicability to a wide range of systems
mainly stems from the nonlinear interaction of exponential-type with nearest-neighbour coupling
among the particles, which reduces to that of coupled harmonic oscillators model in
appropriate limit. The Toda lattice has been used to understand
heat propagation in lattice systems\cite{heat,heat-1}, to model the dynamics
of DNA\cite{dna} and to represent the potential of hydrogen and peptide bonds in the $\alpha$
helix\cite{helix}.  The Toda oscillator has also been used as a simple model of laser
dynamics\cite{ga,ysmj,scfc}. 

There are various generalizations of the Toda lattice where the system is generally
non-integrable. For example, the unequal-mass Toda system shows chaotic behaviour\cite{casati}.
The Toda lattice with inhomogenities in the spring constants\cite{inhomo}, truncated
Toda potential\cite{habib}, dimerized lattice\cite{topo} etc. incorporate realistic physical
situations and exhibit interesting physical behaviours. One of the important generalizations
is dissipative Toda lattice with or without the external driving term\cite{dissi1,dissi2,dissi3,
dissi4,dissi5}. The dynamics of solitons or waves in Toda lattice
in presence of dissipation has been studied leading to decaying as well as oscillatory modes
depending on the specific models. The sustained oscillations and chaotic behaviour are
seen in dissipative Toda lattice with driving term.

Over the past few years, systems with balanced loss and gain have drawn a lot of attention\cite{pkg-review}. One of
the distinguishing characteristics of such a system is that the flow preserves the volume in the position-velocity
state space, although the individual degrees of freedom are subjected to loss or gain\cite{pkg-review}. The system
is non-dissipative and may admit periodic solutions within some regions in the parameter-space. For example,
an experimentally realized ${\cal{PT}}$-symmetric coupled resonators\cite{bpeng} is modeled by
two coupled harmonic oscillators with balanced loss-gain which admits periodic solutions
within suitable ranges of the parameters\cite{ben}. The two-particle rational Calogero model with
balanced loss-gain\cite{ds-pkg}, a chain of coupled oscillators with balanced loss-gain\cite{ben1},
a dimer model of nonlinear oscillators with balanced loss-gain\cite{khare-0} 
also admit periodic solutions. The examples of Hamiltonian system with balanced loss and gain include
systems with nonlinear interaction\cite{ds-pkg, ivb, khare,pkg-ds, ds-pkg1,p6-deb,pkg-pr,pr-pkg}, many-particle
systems\cite{ben1, pkg-ds, ds-pkg1,p6-deb}, systems with space-dependent loss-gain terms\cite{ds-pkg1},
systems with Lorentz interaction\cite{pkg-1} etc., and all of these models admit periodic solutions.
The nonlinear Schr\"odinger equation, nonlinear Dirac Equation and Oligomer have been studied
in presence of balanced loss-gain with important applications in the context of optics, Bose-Einstein
condensate and lattice systems\cite{pkg-review,kono,sg}.

The purpose of this article is to study Toda oscillators with balanced loss and gain. Firstly, we consider
two identical particles in a periodic Toda potential subjected to the balanced loss-gain. The
system is integrable. The integrals of motion consist of the Hamiltonian and a conserved quantity related
to the translation invariance of the system, which may be identified as total generalized momentum.
In the centre of mass coordinate, the system is described in terms of a particle in a harmonic plus hyperbolic
cosine potential. The nature of the effective potential is that of a single-well below certain
critical value of the loss-gain strength. The potential becomes symmetric or asymmetric double-well above
this threshold for vanishing and non-vanishing generalized total momentum, respectively.
Exact analytic solutions may be obtained in some limiting cases in terms of elliptic
functions. However, the model in its full generality does not appear to be amenable to exact,
analytic solutions. The numerical analysis of the system shows existence of periodic solutions
for arbitrary values of the loss-gain strength.

Our main results are based on a system of three identical particles in a periodic Toda potential
with balanced loss-gain and velocity mediated coupling. The loss and gain are balanced between
two particles and the remaining particle is not subjected to any gain or loss. However, all three
particles interact with each other through the Toda potential and velocity mediated coupling.
In general, such velocity mediated coupling appears in the study of many systems including models with
Lorentz interaction, synchronization of different types of oscillators\cite{vsa, ivan,apk} and in the
description of partially ionized plasma\cite{bpm}. The generic system is non-Hamiltonian for arbitrary values
of the strengths of the velocity mediated coupling and loss-gain terms. However, the total generalized
momenta is conserved which is a manifestation of the translation invariance of the system.
The system is Hamiltonian whenever the strength of the loss-gain terms is  two times the strength
of the velocity mediated coupling. We are unable to find any third integral, in addition to the
Hamiltonian and the generalized total momentum, implying that the system may not be integrable.
We study the two-parameter bifurcation diagram numerically which shows that the system is chaotic
for a very large region in the parameter-space. The regular dynamics with periodic solutions
is also observed.  The regular and chaotic dynamics of the system is studied in detail  for the
two limiting cases, (i) vanishing velocity mediated coupling for which the system is non-Hamiltonian 
and (ii) the Hamiltonian system in which the strength of the loss-gain terms is  two times the strength
of the velocity mediated coupling. Theses two cases will be referred to as non-Hamiltonian and Hamiltonian
systems, respectively, throughout this article.

The non-Hamiltonian system is shown to admit periodic solutions around an equilibrium point within
a narrow range of the strength of the loss-gain which is confirmed numerically.
The numerical investigations reveal chaotic behaviour in the system beyond some critical value of the
loss-gain strength. This result is new and deserves some attention. The Toda lattice with identical
particles is an integrable system. The chaotic behaviour is seen for unequal-mass Toda lattice\cite{casati}
and/or in presence of damping and external driving term\cite{dissi1}. The system
under consideration is of identical particles and undriven. Thus, the balanced loss-gain
term is leading to the chaotic behaviour in the system.  The complex dynamical
behavior of  the system is investigated numerically by computing time-series, Poincar$\acute{e}$-sections,
power-spectra, auto-correlation function, bifurcation diagrams and the Lyapunov exponents.

The Hamiltonian system admits regular periodic solution in a wide range of the loss-gain strength
compared to its non-Hamiltonian counterpart. Thus, velocity mediated coupling is responsible for
enhancing the range. This is consistent with the general prescription given in Ref. \cite{pkg-1}
for enhancing the stability region. The system has two integrals of motion, the Hamiltonian and the generalized
total momentum. The numerical analysis shows that the system is chaotic beyond certain ranges of the
loss-gain strength. The systems is investigated numerically in detail by computing time-series,
Poincar$\acute{e}$-sections, power-spectra, auto-correlation function, bifurcation diagrams and
the Lyapunov exponents. The highest value of the Lyapunov exponent is seen to increase with the
increasing values of the loss-gain strength. The quasi-periodic route to chaos is seen numerically.

The plan of the article is the following. We introduce the two-particle periodic Toda system in the
next section. We show that the system is integrable. We obtain regular periodic solution numerically.
In section 3, we introduce the three-particle periodic Toda system with balanced loss-gain and velocity
mediated coupling. The linear stability analysis and the two-parameter bifurcation diagram are presented
at the beginning. The results of numerical investigations for the non-Hamiltonian system are presented
in Sec. 3,1, while that of the Hamiltonian system is described in Sec. 3.2. Finally, the results are
summarized in Sec. 4. 

\section{Periodic Toda Lattic: Two Particles}

The system is defined by the equations of motion,
\bea
&& \ddot{x}_1 +2 \gamma \dot{x}_1 + 2 a \left [ e^{-b (x_2-x_{1})} - e^{-b (x_1-x_{2})}\right ]=0\nonumber \\
&& \ddot{x}_2 - 2 \gamma \dot{x}_2 + 2 a \left [ e^{-b(x_{1}-x_2)} - e^{-b(x_{2}-x_1)} \right ] =0
\label{2TEOM}
\eea
\noindent The loss-gain is balanced since the sum of the coefficients of the loss-gain terms is zero. In
the limit of $b \rightarrow 0, a \rightarrow \infty$ such that $ab\equiv \omega^2$ is finite, the above
system describes a specific case of the coupled oscillators model\cite{ben} in which it is translation
invariant. In this limit, the model is relevant for description of experimentally realized 
${\cal{PT}}$-symmetric coupled resonators\cite{bpeng}. However, for large amplitudes, the nonlinear terms
become relevant and the Toda lattice with balanced loss-gain defines albeit a new system which has not been
considered previously. We define the Hamiltonian of the system as,
\bea
H = 2 \left ( p_1 + \frac{\gamma}{2} x_2 \right )\left ( p_2 - \frac{\gamma}{2} x_1 \right ) +
\frac{a}{b} \left ( e^{b(x_1-x_2)} + e^{b(x_2-x_1)}-2 \right )
\eea
\noindent The canonically conjugate pairs are $(x_1, p_1)$ and $(x_2, p_2)$.
The Hamilton's equation of motion may be used to obtain Eq. (\ref{2TEOM}). The system is integrable
where the second integral of motion in addition to the Hamiltonian has the expression,
\bea
\Pi= \frac{1}{2} \left ( \dot{x_1} + \dot{x}_2 \right ) +\gamma \left (x_1-x_2\right)
\eea
\noindent  It can be checked that $H$ and $\Pi$ are in involution.

We define the centre of mass co-ordinates,
\bea
X=\frac{1}{2} \left (  x_1 -x_2 \right ), \
Y=\frac{1}{2} \left (  x_1 + x_2 \right ),
\eea
\noindent for which the equations of motion described by Eq. (\ref{2TEOM}) reduces to the following equations,
\bea
&& \ddot{Y} + 2 \gamma \dot{X}=0\nonumber \\
&& \ddot{X} + 2 \gamma \dot{Y} + 8 a \sinh (2 b X)=0
\label{cm-cordinate}
\eea
\noindent The trivial translation motion of the centre of mass may be separated by using the
constant of motion $\Pi = \dot{Y} + 2 \gamma X$, which when substituted in the second equation of
(\ref{cm-cordinate}) leads to the decoupled equation,
\bea
\ddot{Z} - 4 \gamma^2 Z + 8 a \sinh \left (2 b \left (Z+\frac{\Pi}{2 \gamma}\right ) \right )=0
\label{Z-eqn}
\eea
\noindent where $Z=X-\frac{\Pi}{2 \gamma}$. The solution for $Y(t)$ may be obtained as,
\bea
Y(t)= \Pi_0 - 2 \gamma \int^t Z(t^{\prime})dt^{\prime}
\label{ref1}
\eea
\noindent where $\Pi_0$ is an integration constant. We choose the constant of motion $\Pi=0$ by imposing
the initial condition $\dot{Y}(0)=-2 \gamma X(0)=-2 \gamma Z(0)$. The initial conditions $X(0)=Z(0)$ and
$\dot{X}(0)=\dot{Z}(0)$ may be chosen depending on the physical requirements. The initial condition
on $Y(0)$ may be chosen by fixing the integration constant $\Pi_0$. The equation for $Z$ with $\Pi=0$
reduces to a simple harmonic oscillator $\ddot{Z} + \Omega^2 Z=0$ in the limit
$b \rightarrow 0, a \rightarrow \infty, ab\equiv \frac{\omega^2}{4}$ with angular frequency
$\Omega^2=4 \left ( \omega^2 -\gamma^2 \right )$. The periodic solutions are obtained for
$- \omega < \gamma <  \omega$. The solutions are unbounded outside this region and are determined
in terms of hyperbolic functions. We may choose small $b \ll 1$ for finite $a$ and express hyperbolic
sine function in Eq. (\ref{Z-eqn}) in powers of $b$. The resulting equation up to the order of $b^3$
with $\Pi=0$ and neglecting all higher order terms describes Duffing oscillator,
\bea
\ddot{Z} + 4 \left ( 4ab -\gamma^2 \right ) Z + \frac{32 a b^3}{3} Z^3=0
\label{ref2}
\eea 
\noindent which is exactly solvable and admits solutions in terms elliptic functions. Unlike the case
of coupled harmonic oscillators, the periodic solutions are obtained even for $\gamma^2 > 4 ab$. We
refer to the Ref. \cite{pkg-ds} for details where solutions of equations similar to Eqs. (\ref{ref1})
and (\ref{ref2}) are discussed. 

\begin{figure}[ht!]
\begin{subfigure}{.5\textwidth}
\centering
\includegraphics[width=.8\linewidth]{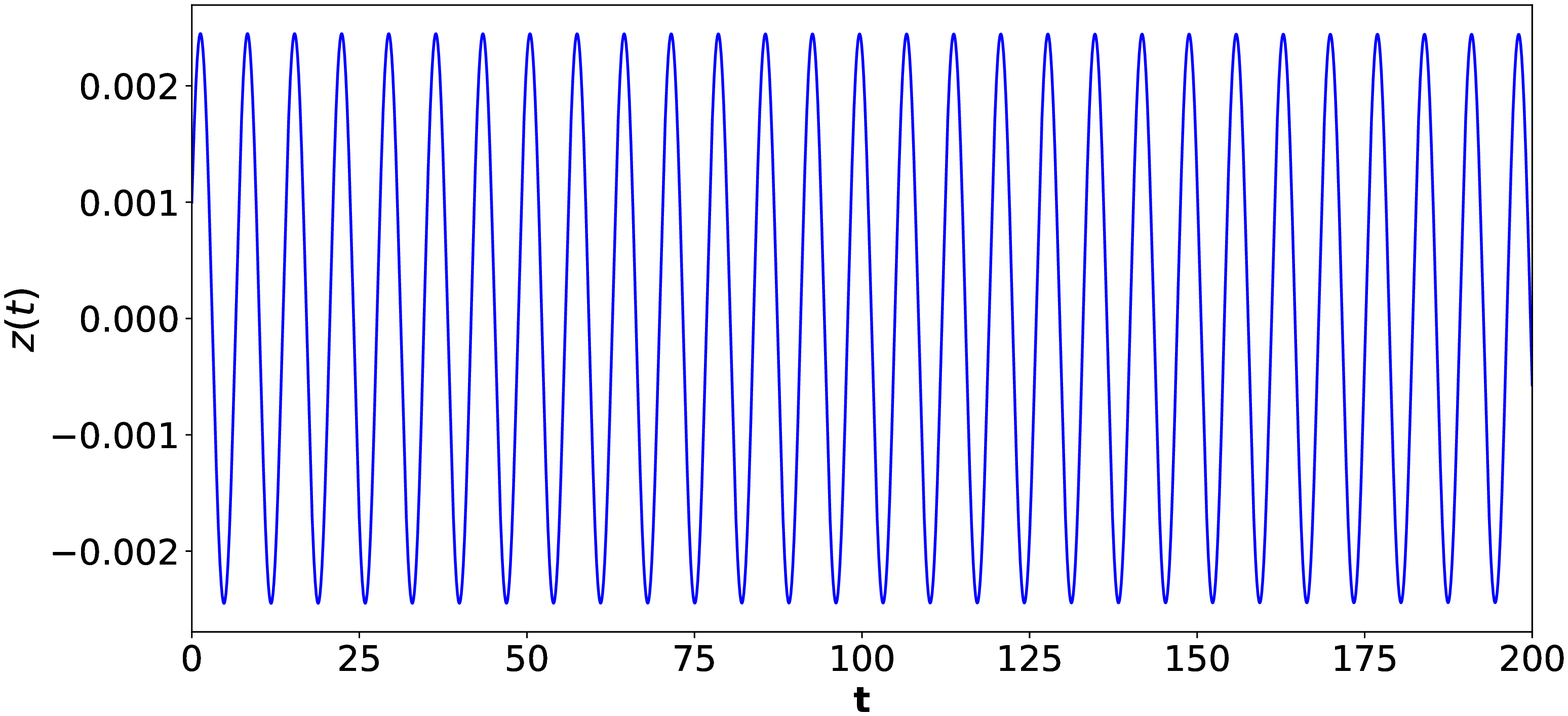} 
\caption{ Regular solution( $\Gamma=0.2, Z_{0}=0.0$).}
\label{Regular solution_z(t)_Gamma_.2_z0_0}
\end{subfigure}%
\begin{subfigure}{.5\textwidth}
\centering
\includegraphics[width=.8\linewidth]{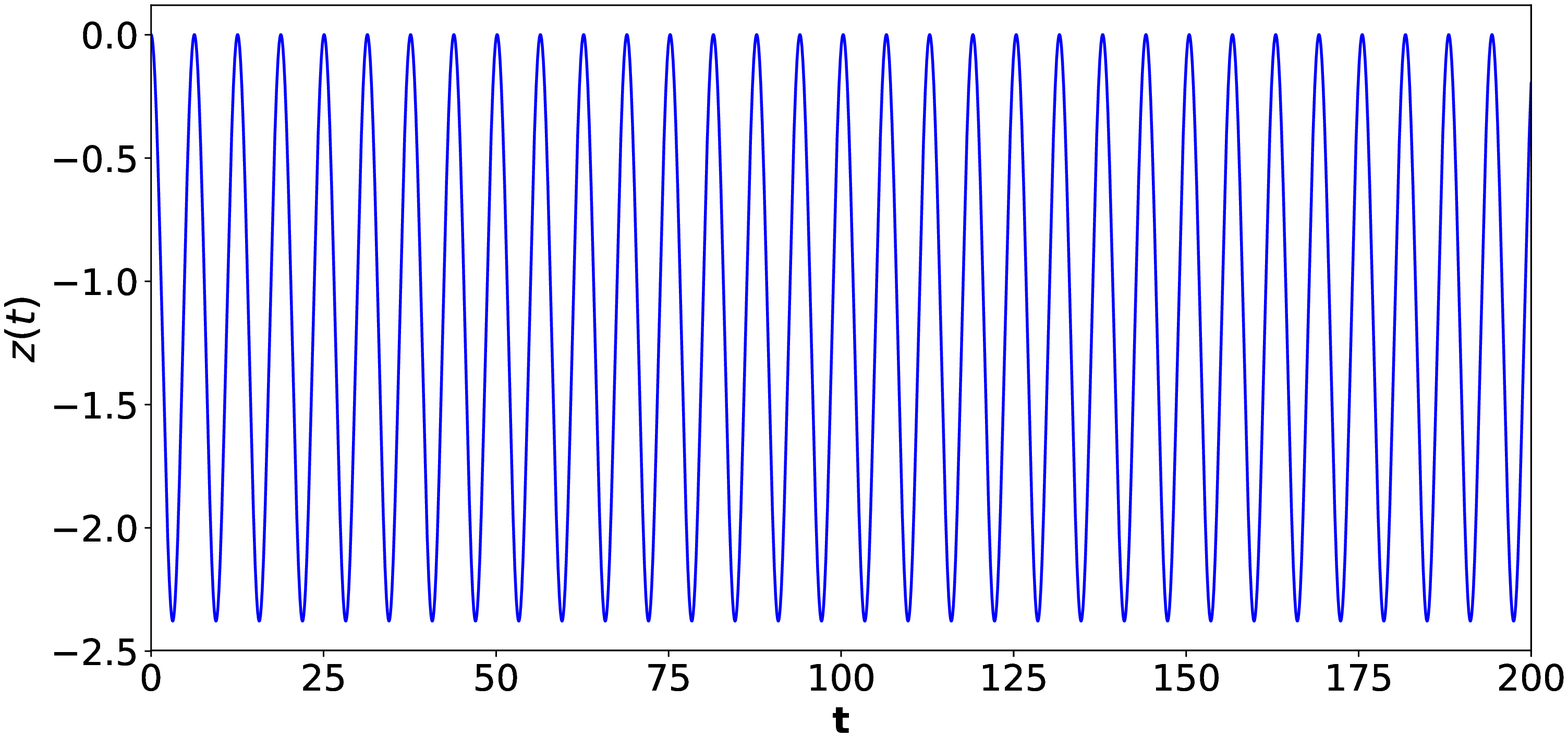}  
\caption{Regular solution( $\Gamma=0.2, Z_{0}=1.0$).}
\label{Regular solution_z(t)_Gamma_.2_z0_1}
\end{subfigure}%
\newline 
\begin{subfigure}{.5\textwidth}
\centering
\includegraphics[width=.8\linewidth]{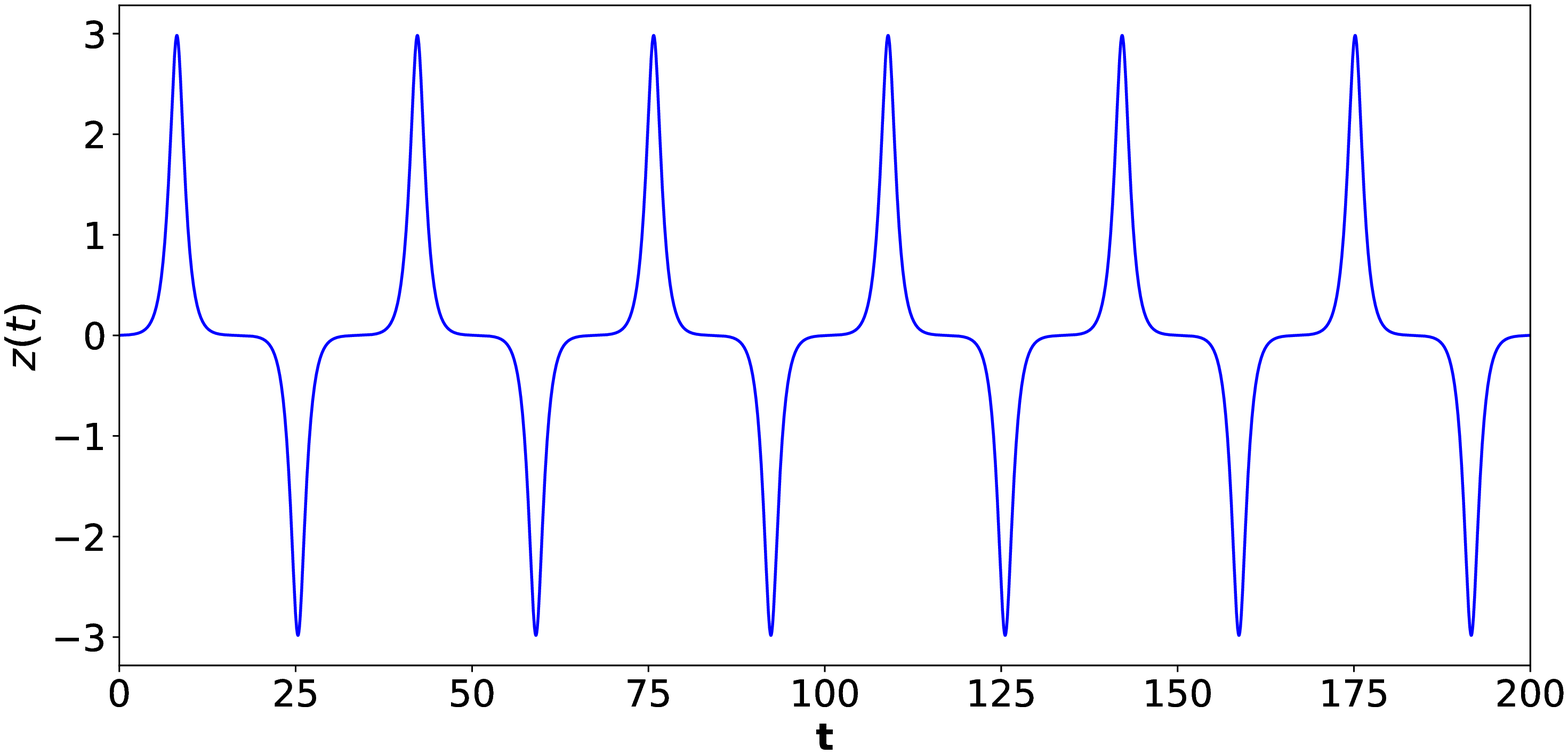} 
\caption{ Regular solution( $\Gamma=2.0, Z_{0}=0.0$)}
\label{Regular solution_z(t)_Gamma_0_z0_1}
\end{subfigure}%
\begin{subfigure}{.5\textwidth}
\centering
\includegraphics[width=.8\linewidth]{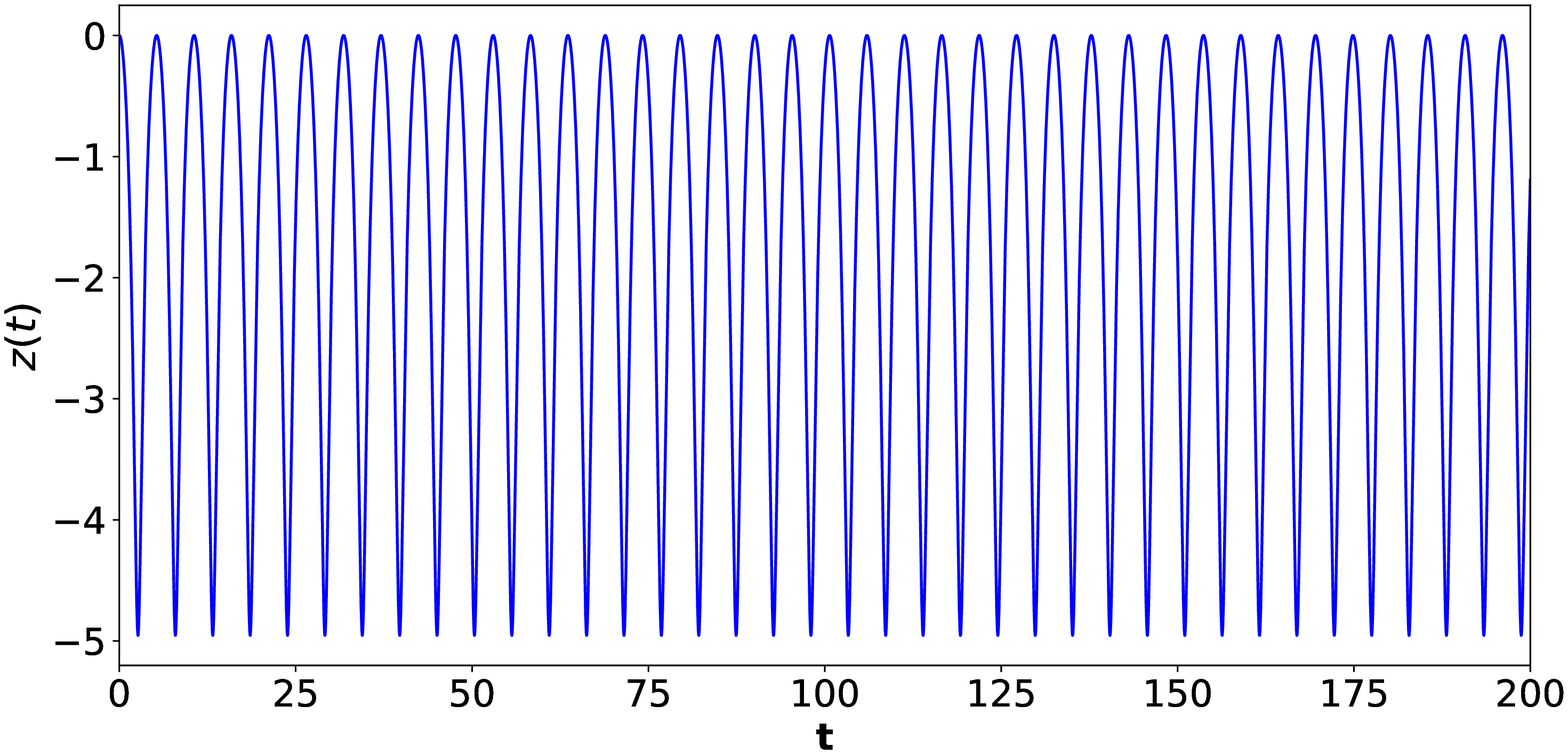}  
\caption{Regular solution($\Gamma=2.0, Z_{0}=1.0$.)}
\label{Regular solution_z(t)_Gamma_2_z0_1}
\end{subfigure}%
\caption{(Color online)  Regular solutions of Eq. (\ref{Z-num}) with the initial conditions $Z(0)=0.001,
 \dot{Z}(0)=0.002$.}
\label{periodic_ solutions-t2}
\end{figure}

It seems that Eq. (\ref{Z-eqn}) with its full generality is not amenable to exact solutions and
we solve it numerically. We employ the scale-transformations $Z \rightarrow \frac{Z}{2b},
t \rightarrow \frac{t}{\sqrt{16ab}}, ab > 0$ in Eq. (\ref{Z-eqn}) which results in the equation.
\bea
\ddot{Z} - \Gamma Z + \sinh \left ( Z + Z_0 \right ) =0
\label{Z-num}
\eea
\noindent where $\Gamma=\frac{\gamma^2}{4 ab}$ and $Z_0=\frac{\Pi b}{\gamma}$. Note that $\Gamma$ is
a measure of the relative strength between the loss-gain and exponential interaction terms. The
constant $Z_0$ may be chosen to be zero by choosing the constant of motion $\Pi=0$.
The Eq. (\ref{Z-num}) describes the motion of a particle in the potential,
\bea
V(Z)=-\frac{\Gamma}{2} Z^2 + \cosh(Z+Z_0)
\eea
\noindent which in the range $\Gamma > \frac{1}{2}$ represents a symmetric double-well potential
for $Z_0=0.0$, while it is an asymmetric double-well potential for $Z_0 \neq 0$. The single-well potential
is obtained for $\Gamma \leq \frac{1}{2}$. The solutions bounded in time exist for all these cases.
We solve Eq. (\ref{Z-num}) for $\Gamma=0.2, 2.0$ for $Z_0=0.0, 1.0$ with various initial conditions which
are plotted in Fig.(\ref{periodic_ solutions-t2}). For the case of double-well potential, the oscillation
of the particle between the two wells can be seen for suitable initial conditions.

\section{Periodic Toda Lattice: Three particles}

The system is described by the equations of motion,
\bea
&& \ddot{q}_1 + \gamma_{1} \dot{q}_1 +g \left ( \dot{q}_2 - \dot{q}_3\right )
-a \left [ e^{b\left(q_2-q_1\right )} -e^{b\left(q_1-q_3\right)}\right]=0\nonumber \\
&& \ddot{q}_2 + g \left ( \dot{q}_1- \dot{q}_3 \right )
-a \left [ e^{b\left(q_3-q_2\right )} -e^{b\left(q_2-q_1\right)}\right]=0\nonumber \\
&& \ddot{q}_3 - \gamma_{1} \dot{q}_3 + g \left ( \dot{q}_1 - \dot{q}_2 \right )
-a \left [ e^{b\left(q_1-q_3\right )} -e^{b\left(q_3-q_2\right)}\right]=0
\label{toda_eqn}
\eea
\noindent  The parameters $\gamma_1$ and $g$ are related to the strengths of the constant
loss-gain and velocity mediated coupling among the particles, respectively.
The parameters $a$ and $b$ are related to the stiffness or the spring-constant of the nonlinear
interaction. The loss-gain is balanced between the first and the third particle with their
coordinates denoted by $q_1$ and $q_3$, respectively. The second particle with the coordinate
$q_2$ is not subjected to any gain or loss. However, each particle is interacting with the other
two particles through the exponential interaction as well as velocity-mediated coupling.
The Eq. (\ref{toda_eqn}) defines a system of balanced loss and gain in the sense that the
flow in the position-velocity state space preserves the volume, although individual degrees
of freedom are subjected to gain or loss. It is possible to formulate the problem wherein
the $i^{th}$ particle is subjected to loss-gain with the strength $\gamma_i$ such that
$\sum_{i=1}^3 \gamma_i=0$. We have chosen a specific case with $\gamma_3=-\gamma_1,
\gamma_2=0$. The qualitative features are same for both these formulations. The system reduces to
a chain of coupled oscillators with balanced loss-gain in the limit $g=0, a \rightarrow \infty,
b \rightarrow 0$ such that  $ab=\omega^2$, which is a variant of the model considered in
Ref. \cite{ben1}. For finite $a$ and $b \ll 1$, the exponential potential may be expanded in
power series and truncated at a desired order in powers of $b$ leading to some effective potential.
For example, if the series is truncated at $O(b^3)$ and the centre of mass is separated,
the H{\'e}non-Heiles system with balanced loss-gain is produced for $g=0$.

The system is translation
invariant and the corresponding conserved quantity $\Pi^{(3)}$ has the expression,
\bea
\Pi^{(3)}=\dot{q}_1+\dot{q}_2+\dot{q}_3 + \left(\gamma_{1}+2 g\right)
\left ( q_1 -q_3 \right )
\eea
\noindent The integral of motion $\Pi^{(3)}$ may be identified as the generalized total momentum
which reduces to the total momentum $p_1+p_2+p_3$ for $g=-\frac{1}{2} \gamma_1$.
It may be noted that the case of Toda lattice, i.e. $\gamma_1=g=0$ is included in the
above reduction. It seems that the system is non-Hamiltonian for the generic values of
$\gamma_1$ and $g$. The system is Hamiltonian for $g=\frac{1}{2} \gamma_1 \equiv
\gamma$:
\bea
H=\Pi_1 \Pi_2 + \Pi_1 \Pi_3 + \Pi_2 \Pi_3 + \frac{a}{b} \sum_{i=1}^3
\left [ 1 - e^{b\left ( q_{n+1} -q_n \right ) } \right ]
\eea
\noindent where the generalized momenta $\Pi_i$ are defined as,
\bea
\Pi_1=p_1 +\frac{\gamma}{4} \left ( q_2+q_3 \right ),\
\Pi_2=p_2 -\frac{\gamma}{4} \left ( q_1-q_3 \right ),\
\Pi_3=p_3 -\frac{\gamma}{4} \left ( q_1+q_2 \right ).
\eea
\noindent We have identified  $q_{4} \equiv q_1$ in the above expression.
The integral of motion takes the form $\Pi^{(3)}=\dot{q}_1+\dot{q}_2+\dot{q}_3 +
2 \gamma \left ( q_1 -q_3 \right )$ and is in involution with $H$.  We have found two
integrals of motion for a system with three degrees of freedom. The Liouville integrability
requires a third integral of motion which is in involution with both $\Pi^{(3)}$ and $H$.
However, a numerical analysis of the system reveals chaotic behaviour within a large range
of the parameter.

The independent scales in the system may be fixed by employing the transformations,
$t \rightarrow \frac{1}{\sqrt{ab}} t, \ q_{i} \rightarrow \frac{1}{b} q_{i}, i=1,2,3$.
where $ab > 0$. The equations of motion in terms of the scaled variables are, 
\bea
&& \ddot{q}_1 + \Gamma_{1} \dot{q}_1 +\Gamma_{2} \left ( \dot{q}_2 - \dot{q}_3\right )
- \left [ e^{\left(q_2-q_1\right )} -e^{\left(q_1-q_3\right)}\right]=0\nonumber \\
&& \ddot{q}_2 + \Gamma_{2} \left ( \dot{q}_1- \dot{q}_3 \right )
- \left [ e^{\left(q_3-q_2\right )} -e^{\left(q_2-q_1\right)}\right]=0\nonumber \\
&& \ddot{q}_3 - \Gamma_{1}\dot{q}_3 + \Gamma_{2} \left ( \dot{q}_1 - \dot{q}_2 \right )
- \left [ e^{\left(q_1-q_3\right )} -e^{\left(q_3-q_2\right)}\right]=0
\label{toda_eqn1}
\eea
\noindent where $\Gamma_1=\frac{\gamma_1}{\sqrt{ab}}$ and $\Gamma_2=\frac{g}{\sqrt{ab}}$.
Unlike the Toda lattice, the parameters $a$ and $b$ can not be removed completely unless
$\gamma_1= g = 0$, i.e. vanishing loss-gain as well as velocity-mediated coupling.
The total number of independent parameters is reduced from four to two. The parameter
$\Gamma_1$ measures the relative strength between the gain-loss term and stiffness of the
nonlinear spring, while $\Gamma_2$ measures the relative strength between the velocity mediated
coupling and stiffness of the nonlinear spring. We will present linear stability analysis and
bifurcation diagram of the system for generic $\Gamma_1$ and $\Gamma_2$. The regular and chaotic
dynamics of the system will be investigated in detail for the two cases (i) $\Gamma_2=0$ 
and (ii) $\Gamma_2=\frac{1}{2} \Gamma_1$. The first case corresponds to Toda lattice with balanced
loss-gain and without any velocity mediated coupling among the particles. The system appears
to be non-Hamiltonian. The second case corresponds to a Hamiltonian system of Toda lattice
with balanced loss-gain and velocity mediated coupling.

The system is translation invariant and we use the Jacobi coordinates to eliminate the centre of
mass motion:
\bea
Q_1=\frac{1}{\sqrt{2}} \left ( q_1-q_2\right ),
Q_2= \frac{1}{\sqrt{6}} \left ( q_1+q_2 - 2 q_3 \right ), \ Q_3=\frac{1}{\sqrt{3}} \left (
q_1+q_2+q_3 \right )
\label{coordinate_center_of_mass}
\eea
\noindent The center of mass coordinate $Q_3$ completely decouples from the equations described by $Q_1$
and $Q_2$ after the substitution of expression of $\dot{Q}_3$ given by the above equation.
The equations for $Q_1$ and $Q_2$ in terms of the scaled variables
$Q_1 \rightarrow \sqrt{2} Q_1, Q_2 \rightarrow \sqrt{\frac{2}{3}} Q_2,
Q_3\rightarrow \sqrt{\frac{2}{3}} Q_3, t \rightarrow {\sqrt{2}} t $ read:
\bea
\ddot{Q}_1 & = & - \frac{\Gamma_{1}-2\Gamma_{2}}{\sqrt{2}}\dot{Q}_{1}-\frac{\Gamma_{1}}{3\sqrt{2}}\dot{Q}_2 -\frac{\Gamma_{1}}{\sqrt{3}}\Pi^{(3)}
 +  \frac{\Gamma_{1}\left(\Gamma_{1}+2\Gamma_{2}\right)}{3 } \left ( Q_1 + Q_2  \right )\nonumber \\
& + & {2} e^{-{2} Q_1} - e^{ \left ( Q_1 +
 Q_2 \right) }  -e^{ \left ( Q_1- Q_2 \right) }\nonumber \\
\ddot{Q}_2 & = & - \frac{\Gamma_{1}-4\Gamma_{2}}{\sqrt{2}}\dot{Q}_{1}+ \frac{\Gamma_{1}-2\Gamma_{2}}{\sqrt{2}}\dot{Q}_{2}-{\sqrt{3}\Gamma_{1}}\Pi^{(3)}
+ {\Gamma_{1}\left(\Gamma_{1}+2\Gamma_{2}\right)} \left ( Q_1 + Q_2  \right )\nonumber \\
& - & {{3}} e^{ \left ( Q_1 + Q_2 \right) }  + {3} e^{ \left ( Q_1- Q_2 \right) }
\label{eqnn_center_of_mass}
\eea
\noindent and the constant of motion $\Pi^{(3)}$ becomes
$\Pi^{(3)}=\frac{1}{\sqrt{3}} \left [ \dot{Q}_3 +\left(\Gamma_{1}+2\Gamma_{2}\right)\left (Q_1 + Q_2 \right ) \right ]$.
\noindent It may be noted that the balanced loss-gain is present in the reduced system described by $Q_1$
and $Q_2$, since the coefficient of $\dot{Q}_1$ in the first equation is exactly opposite to
the coefficient of $\dot{Q}_2$ in the second equation. The loss-gain terms are absent in the Hamiltonian
system characterized by $\Gamma_2=\frac{1}{2} \Gamma_1$. The velocity mediated coupling between
the two modes are present in the reduced system. The harmonic terms vanish for $\Gamma_2=
-\frac{1}{2} \Gamma_1$ for which the total momentum is a conserved quantity, i.e. $\Pi^{(3)}=
\frac{\dot{Q}_3}{\sqrt{3}}$. We choose the constant $\Pi^{(3)}=0$ in subsequent calculations
to eliminate the
trivial translation motion. Various effective systems may be obtained by expanding the exponential
terms and truncating the series at a desired order. For example, a variant of the H{\'e}non-Heiles
system with balanced loss-gain may be produced from Eqn. (\ref{eqnn_center_of_mass}) by expanding
the exponential terms in power series and truncated at the second order. The harmonic terms depend
on $\Gamma_1, \Gamma_2$ and it seems, unlike the standard  H{\'e}non-Heiles system, it is
non-Hamiltonian for $\Gamma_1 \neq \frac{1}{2} \Gamma_2$.

\begin{figure}
\begin{subfigure}{.5\textwidth}
\centering
\includegraphics[width=.8\linewidth]{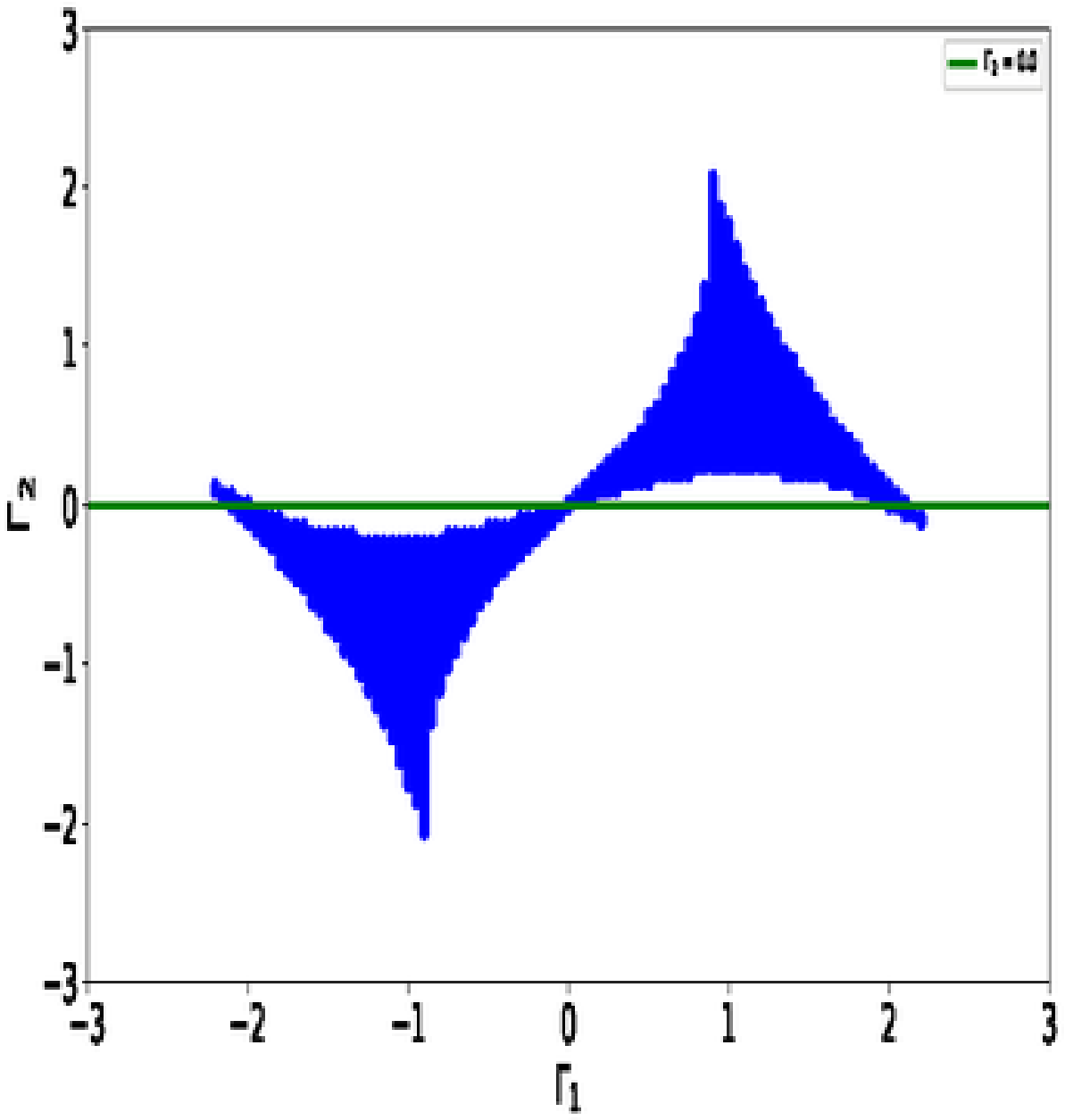} 
\caption {Region of periodic solutions in $\Gamma_1-\Gamma_2$ space\\ is denoted by
	blue color}
\label{Region-period}
\end{subfigure}%
\begin{subfigure}{.5\textwidth}
\centering
\includegraphics[width=.8\linewidth]{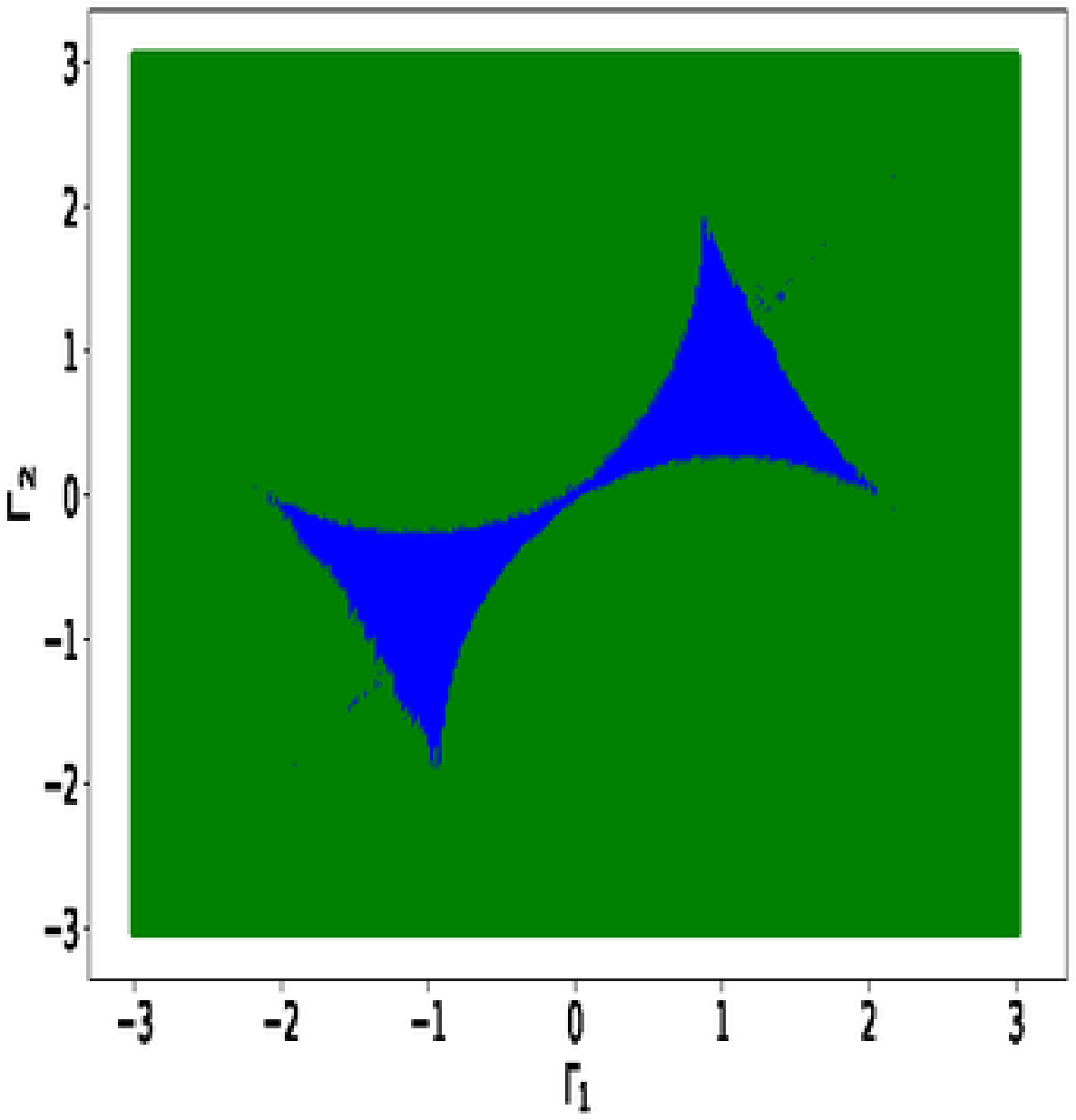} 
\caption{ Bifurcation diagram in $\Gamma_1-\Gamma_2$ space. The non-chaotic region is colored in blue}
\label{bifurcation-2p}
\end{subfigure}%
\caption{(Color online) Region of periodic solution predicted by linear stability analysis and
bifurcation diagram in the `$\Gamma_1-\Gamma_2$' parameter space }
\end{figure}

The equilibrium points and their stability may be analyzed by employing
standard techniques. We denote $\dot{Q}_1=Q_3, \dot{Q}_2=Q_4, Q^T \equiv(Q_1, Q_2, Q_3, Q_4)$, where
the superscript $^T$ denotes transpose. The point $Q=0$ in the position-velocity state-space is an
equilibrium point. Considering fluctuations $\xi$ around $Q=0$ and keeping only the linear terms
in Eq. (\ref{eqnn_center_of_mass}), we obtain $\dot{\xi}=M \xi$, where 
\bea
M=\begin{pmatrix}
0 & 0 & 1 & 0 \\
0 & 0 & 0 & 1 \\
\frac{\Gamma_1(\Gamma_1+2\Gamma_2)}{3}-6 &  \frac{\Gamma_1(\Gamma_1+2\Gamma_2)}{3} &
 \frac{-\Gamma_1+2\Gamma_2}{\sqrt{2}} & -\frac{\Gamma_1}{3\sqrt{2}} \\
\Gamma_1(\Gamma_1+2\Gamma_2) & {\Gamma_1(\Gamma_1+2\Gamma_2)}-6 & \frac{-\Gamma_1 +
4 \Gamma_2}{\sqrt{2}} & \frac{\Gamma_1-2 \Gamma_2}{\sqrt{2}} \\
\end{pmatrix}
\eea
\noindent having eigenvalues $\lambda_{\pm}, -\lambda_{\pm}$:
\bea
\lambda_{\pm} = \left ( \Gamma_1^2 + \Gamma_2^2 -6 \pm \left [ \left (\Gamma_1^2+\Gamma_2^2 \right )^2 +
4 \Gamma_2^2 -4 \left (\Gamma_1 -2 \Gamma_2 \right )^2 \right ]^{\frac{1}{2}} \right )^{\frac{1}{2}}
\eea
\noindent In general, the eigenvalues are complex. The periodic solutions are obtained when all four
eigenvalues are purely imaginary. For the special case of $\Gamma_2=0$, i.e.
vanishing velocity mediated coupling, periodic solutions are obtained in a very narrow range,
\bea
2 \leq {\vert \Gamma_1 \vert} < \frac{3}{\sqrt{2}} \approx 2.1213
\label{grange}
\eea
\noindent and also for $\Gamma_1=0$, i.e. the vanishing loss-gain terms. The range is
enhanced significantly for the Hamiltonian system, i.e. $\Gamma_2=\frac{1}{2} \Gamma_1$,
due to the presence of the velocity mediated coupling, and the periodic solutions are obtained
in the range,
\bea
- \frac{3}{2} < \Gamma_1 < \frac{3}{2} 
\label{grange-1}
\eea
\noindent There are no periodic solutions for vanishing loss-gain and non-vanishing velocity
mediated coupling, i.e. $\Gamma_1=0, \Gamma_2 \neq 0$. The region in the `$\Gamma_1-\Gamma_2$'
parameter-space in which periodic solutions exist is determined numerically and plotted in Fig. 
\ref{Region-period}.
The system defined by Eq. (\ref{eqnn_center_of_mass}) also admits equilibrium points
$Q=(q_0, 3 q_0, 0, 0)$ for an arbitrary real constant $q_0$, whenever $q_0 \Gamma_1(\Gamma_1+2\Gamma_2)
=\frac{3}{2}  e^{ q_0} \sinh(3 q_0)$ for various combinations of $\Gamma_1$ and $\Gamma_2$. However,
these points are hyperbolic in nature and no stable solutions can be obtained. These equilibrium points
are spurious with respect to the original three-particle problem, since no such equilibrium point
exists in terms of $q_1, q_2, q_3$.

The system described by Eqn. (\ref{eqnn_center_of_mass} ) contains two independent parameters $\Gamma_1$
and $\Gamma_2$. The bifurcation diagram in the `$\Gamma_1-\Gamma_2$' parameter-space
is  shown in Fig. \ref{bifurcation-2p}. The bifurcation diagram is drawn with the initial
values of the dynamical variables chosen around the equilibrium point $Q=0$.
The regular periodic solutions are obtained within the region colored in blue, while the chaotic
region is colored in green. The chaotic behaviour for $\Gamma_2=0$ is seen for all values of $\Gamma_1$
excluding the range $2 \leq {\vert \Gamma_1 \vert} < \frac{3}{\sqrt{2}}$ and the point
$\Gamma_1=0$.
It is known that the equal-mass periodic Toda lattice without any external driving term is not
chaotic. The system under consideration for
\begin{figure}[ht!]
\begin{subfigure}{.5\textwidth}
\centering
\includegraphics[width=.8\linewidth]{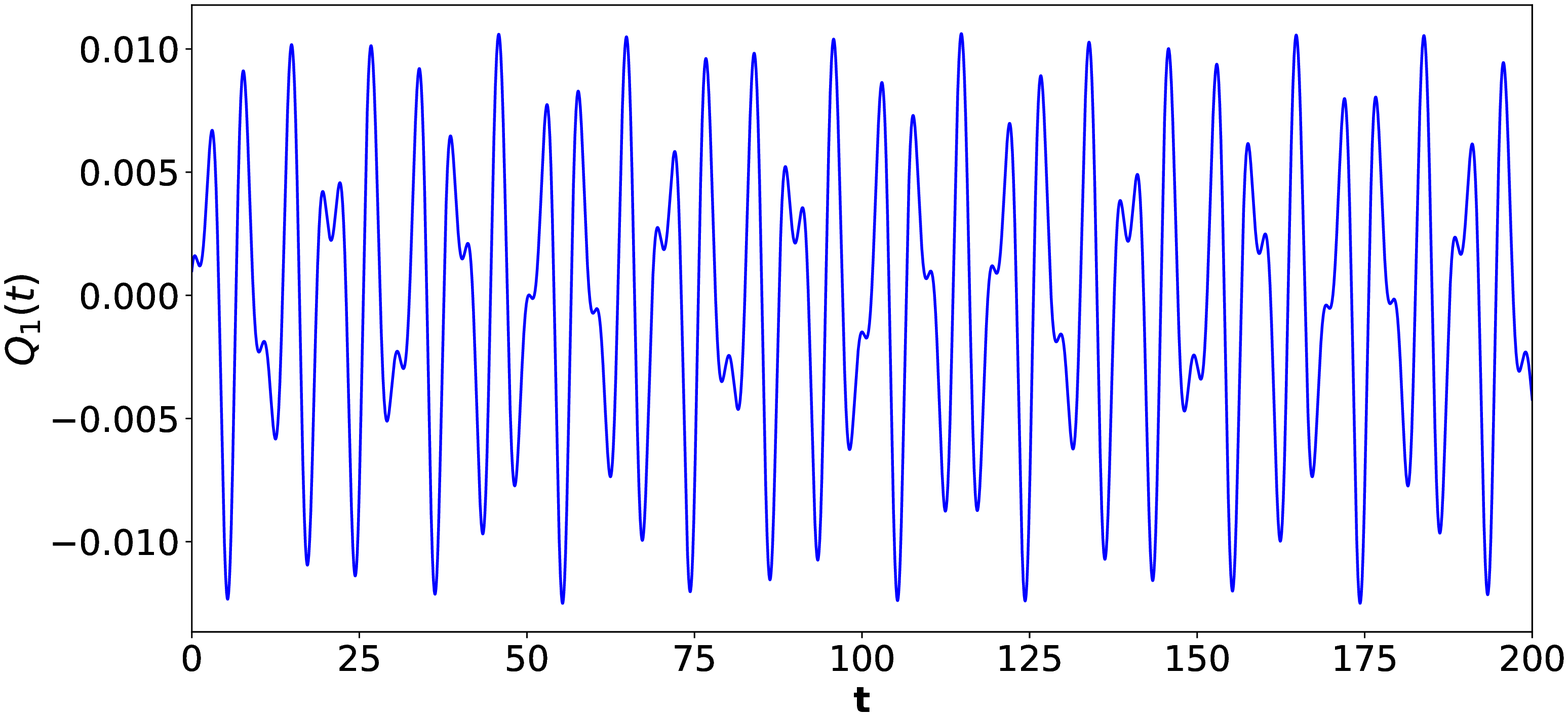} 
\caption{ Regular solution( $\Gamma_1=2.04, \Gamma_2=0$).}
\label{Regular solution_Q1_gamma_+ve}
\end{subfigure}%
\begin{subfigure}{.5\textwidth}
\centering
\includegraphics[width=.8\linewidth]{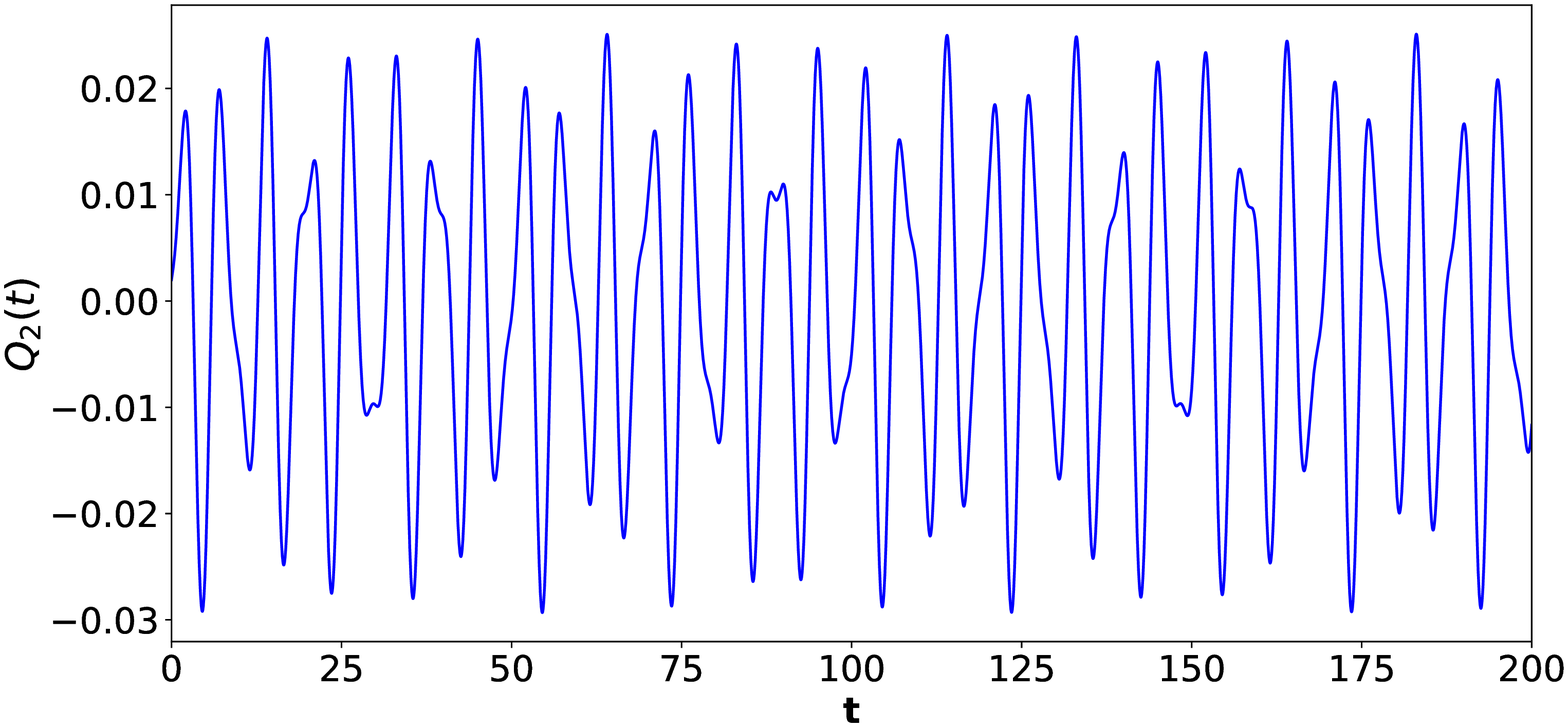} 
\caption{ Regular solution( $\Gamma_1=2.04, \Gamma_2=0$)}
\label{Regular solution_Q2_gamma_+ve}
\end{subfigure}%
\caption{(Color online)  Regular solutions of the non-Hamiltonian Toda system in the vicinity of the point
$Q=0$ with the initial conditions $Q_{1}(0)=0.001, Q_{2}(0)=0.002, \dot{ Q_{1}}=0.003, \dot{Q_{2}}=0.004$.}
\label{periodic_ solutions}
\end{figure}
\begin{figure}[ht!]
\begin{subfigure}{.5\textwidth}
\centering
\includegraphics[width=.8\linewidth]{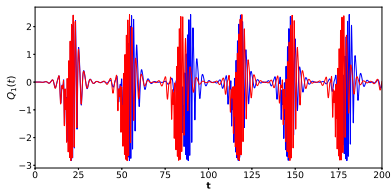} 
\caption{ Chaotic solution( $\Gamma_1=1.85, \Gamma_2=0$).}
\label{Chaotic_solution_Q1_gamma_+ve}
\end{subfigure}%
\begin{subfigure}{.5\textwidth}
\centering
\includegraphics[width=.8\linewidth]{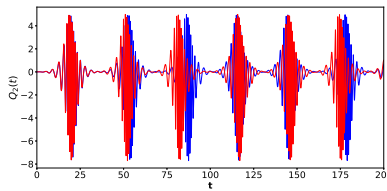} 
\caption{ Chaotic solution( $\Gamma_1=1.85, \Gamma_2=0$)}
\label{Chaotic_solution_Q2_gamma_+ve}
\end{subfigure}%
\caption{(Color online) Chaotic solutions of the non-Hamiltonian Toda system with two sets of initial
conditions (a)$Q_{1}(0)=0.001, Q_{2}(0)=0.002, \dot{ Q_{1}}=0.003,\dot{Q_{2}}=0.004$ (blue color) and 
(b)$Q_{1}(0)=0.001, Q_{2}(0)=0.002, \dot{ Q_{1}}=0.003,\dot{Q_{2}}=0.0045$(red color).}
\label{chaotic solutions}
\end{figure}
$\Gamma_2=0, \Gamma_1 \neq 0$ deals with equal-mass
Toda lattice with balanced loss-gain and without any driving term, yet it shows
chaotic behaviour. Thus, the chaotic behaviour is induced by the loss-gain. The Hamiltonian
system, i.e. $\Gamma_2 =\frac{\Gamma_1}{2}$  is chaotic for ${\vert \Gamma_1 \vert} > \frac{3}{2}$. 
This provides an example of Hamiltonian chaos in Toda lattice with balanced loss-gain. Both the
limiting cases discussed above contain regular as well as chaotic dynamics in the specified regions
of the parameter-space. However, for $\Gamma_1=0, \Gamma_2 \neq 0$, the system is chaotic for any
non-zero values of the velocity-mediated coupling. It may be noted that although the loss-gain
terms are absent in Eq. (\ref{toda_eqn1}) for $\Gamma_1=0, \Gamma_2 \neq0$, it appears
in Eqn. (\ref{eqnn_center_of_mass}). We study the regular and the chaotic dynamics of the system
is some detail for the two cases, (i)$\Gamma_2=0, \Gamma_1 \neq 0$ and (ii) $\Gamma_2 =\frac{\Gamma_1}{2}
\neq 0$, which are referred to as non-Hamiltonian and Hamiltonian systems, respectively. 

\subsection{Non-Hamiltonian System: Numerical results}

\begin{figure}
\begin{subfigure}{.5\textwidth}
\centering
\includegraphics[width=.8\linewidth]{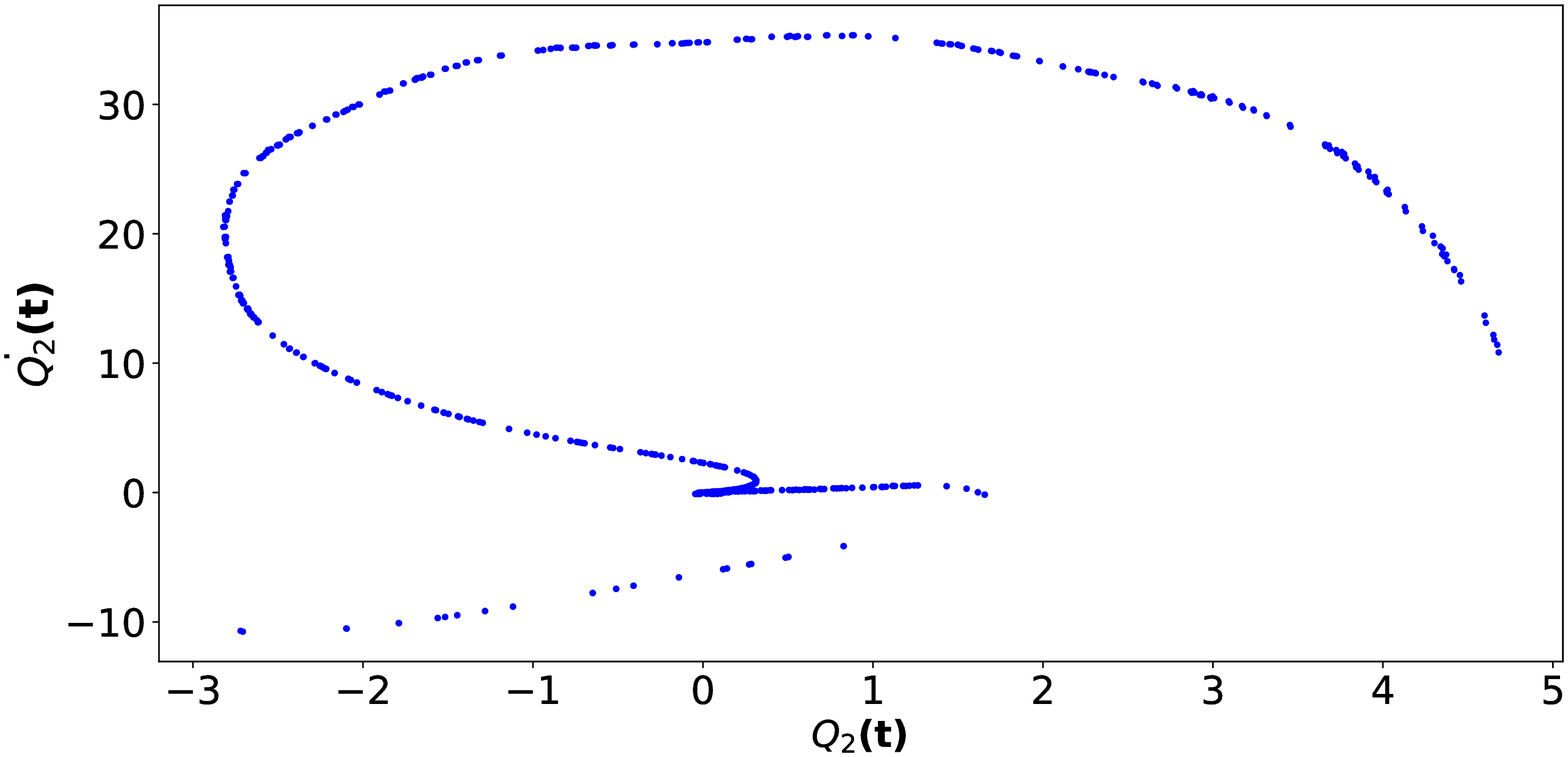}
\caption{Poincar$\acute{e}$ section: $\dot{Q_{2}}(t)$ VS. $Q_{2}(t)$ plot}
\label{multi-poincare-gamma}
\end{subfigure}%
\begin{subfigure}{.5\textwidth}
\centering
\includegraphics[width=.8\linewidth]{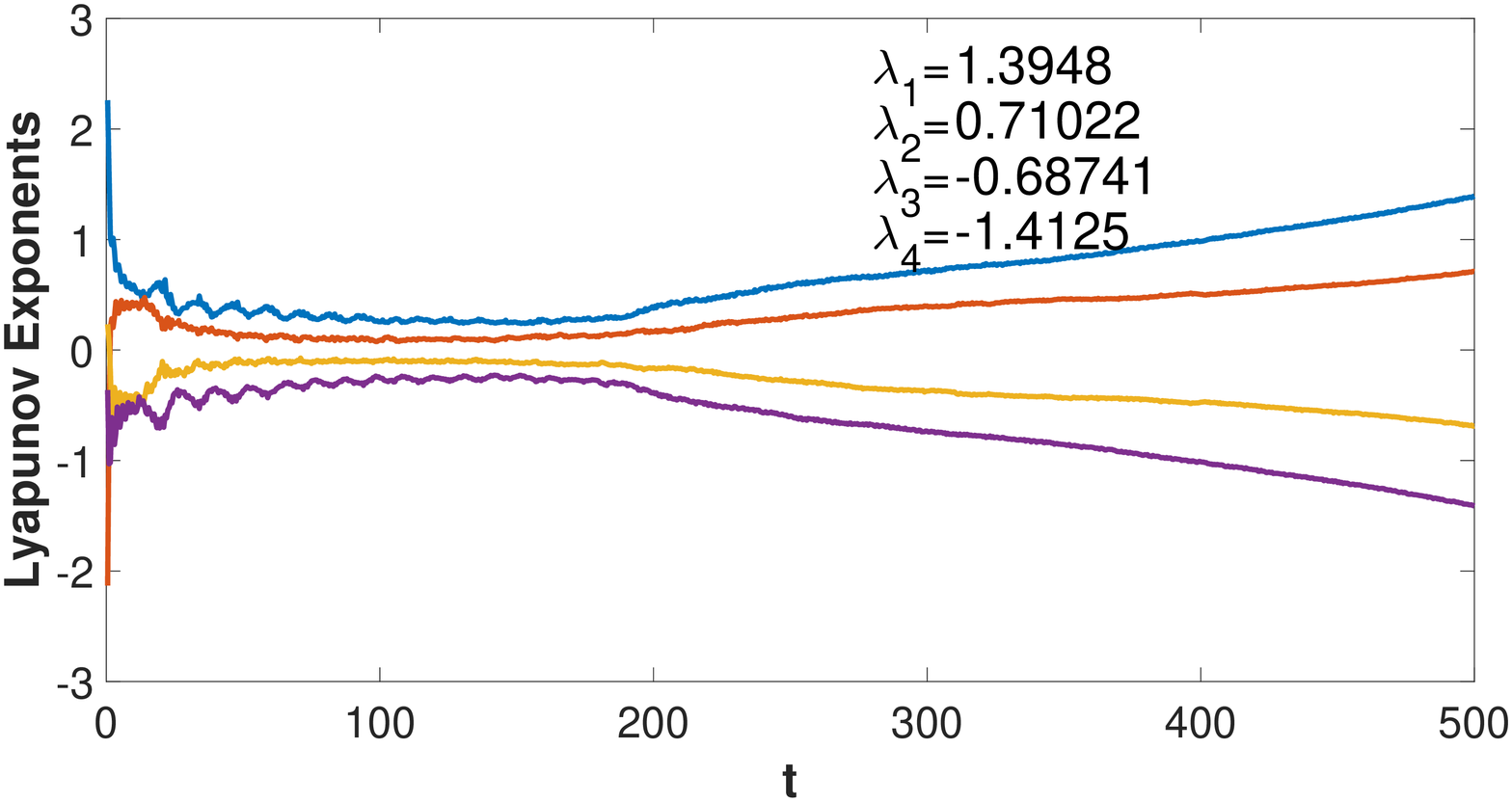}
\caption{Lyapunov exponents}
\label{multi-lyap_gamma_+ve}
\end{subfigure}%
\newline
\begin{subfigure}{.5\textwidth}
\centering
\includegraphics[width=.8\linewidth]{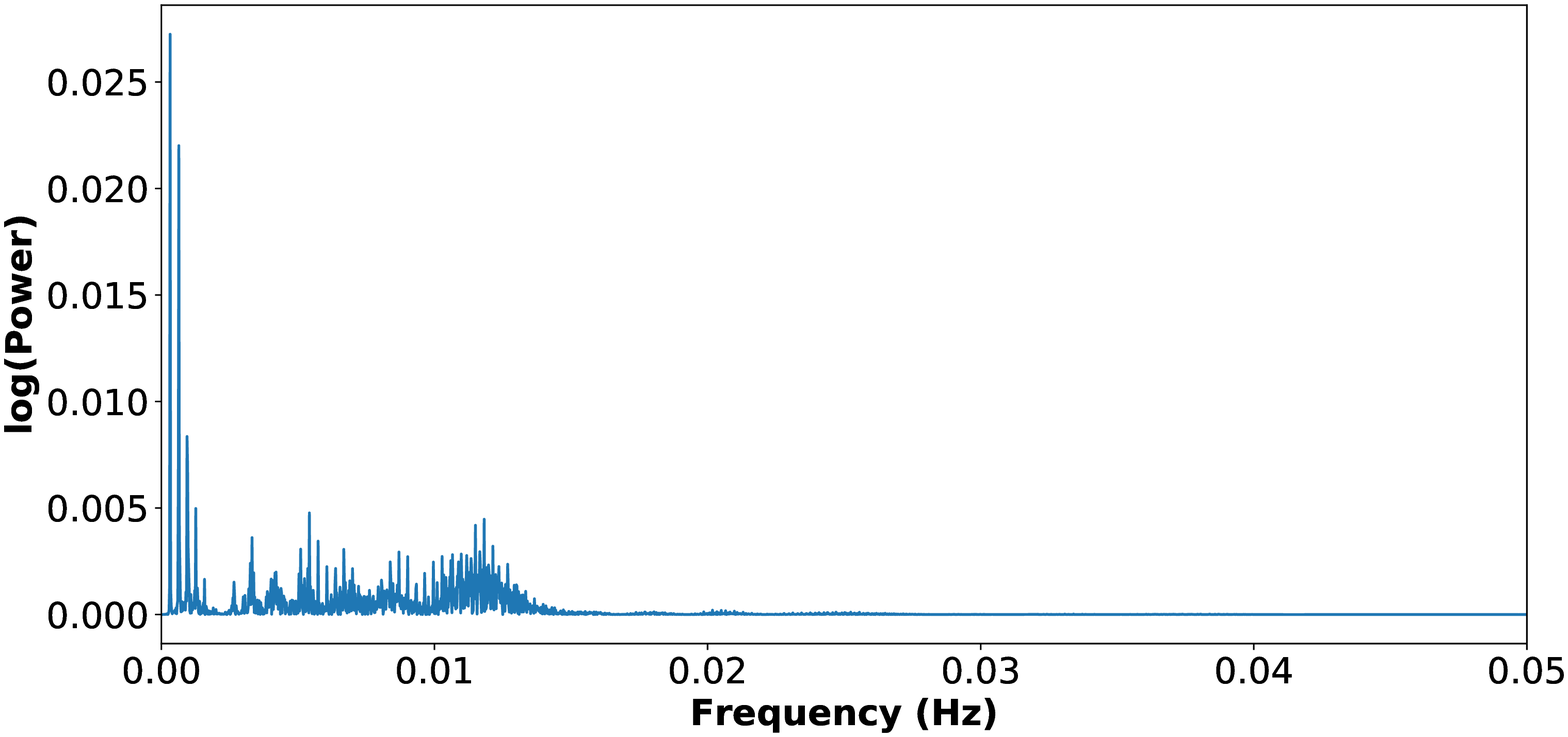}
\caption{Powerspectra  of $q_{1}(t)$}
\label{toda-power_Q1_gamma_+ve}
\end{subfigure}%
\begin{subfigure}{.5\textwidth}
\centering
\includegraphics[width=.8\linewidth]{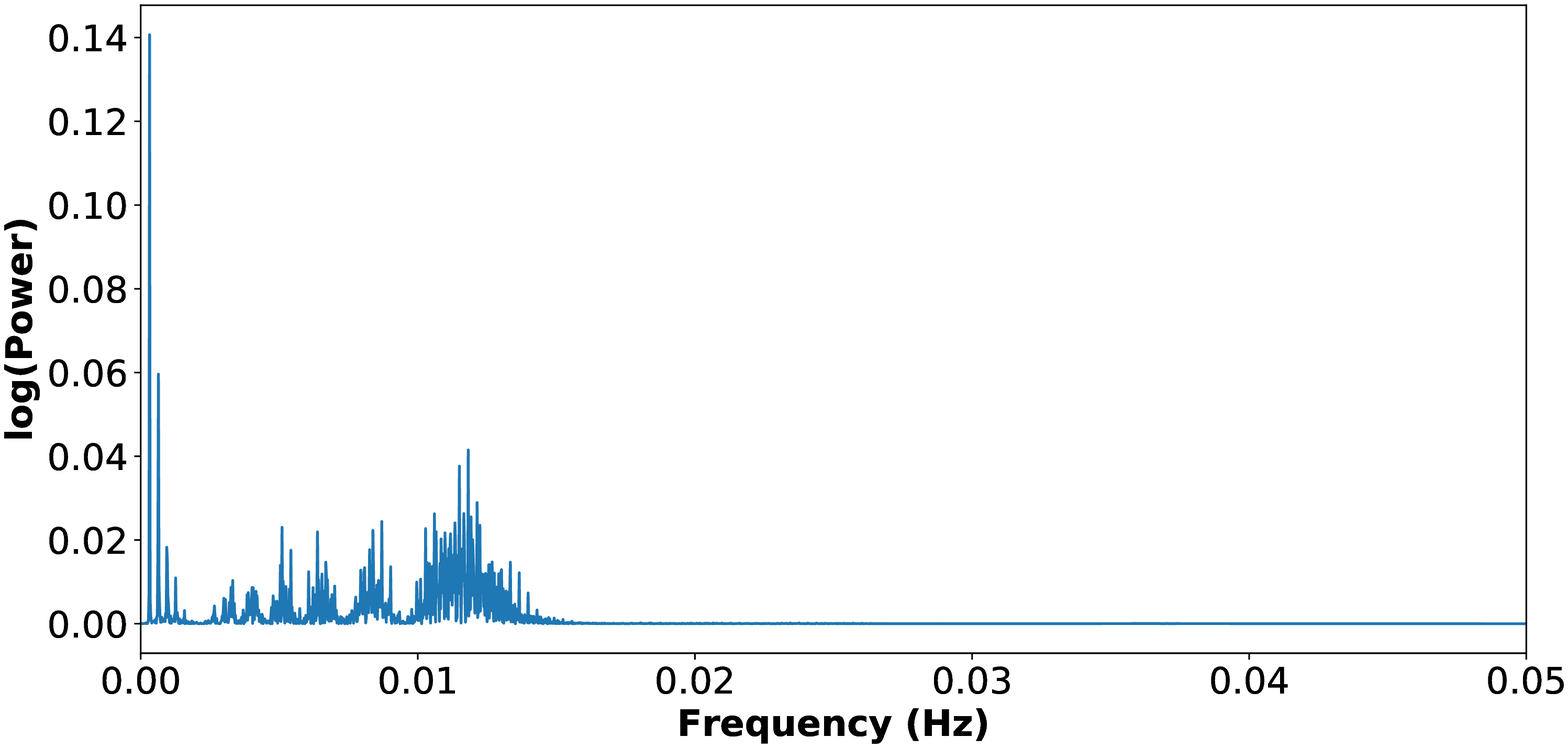}
\caption{Powerspectra  of $q_{2}(t)$}
\label{toda-power_Q2_gamma_+ve}
\end{subfigure}%
\newline
\begin{subfigure}{.5\textwidth}
\centering
\includegraphics[width=.8\linewidth]{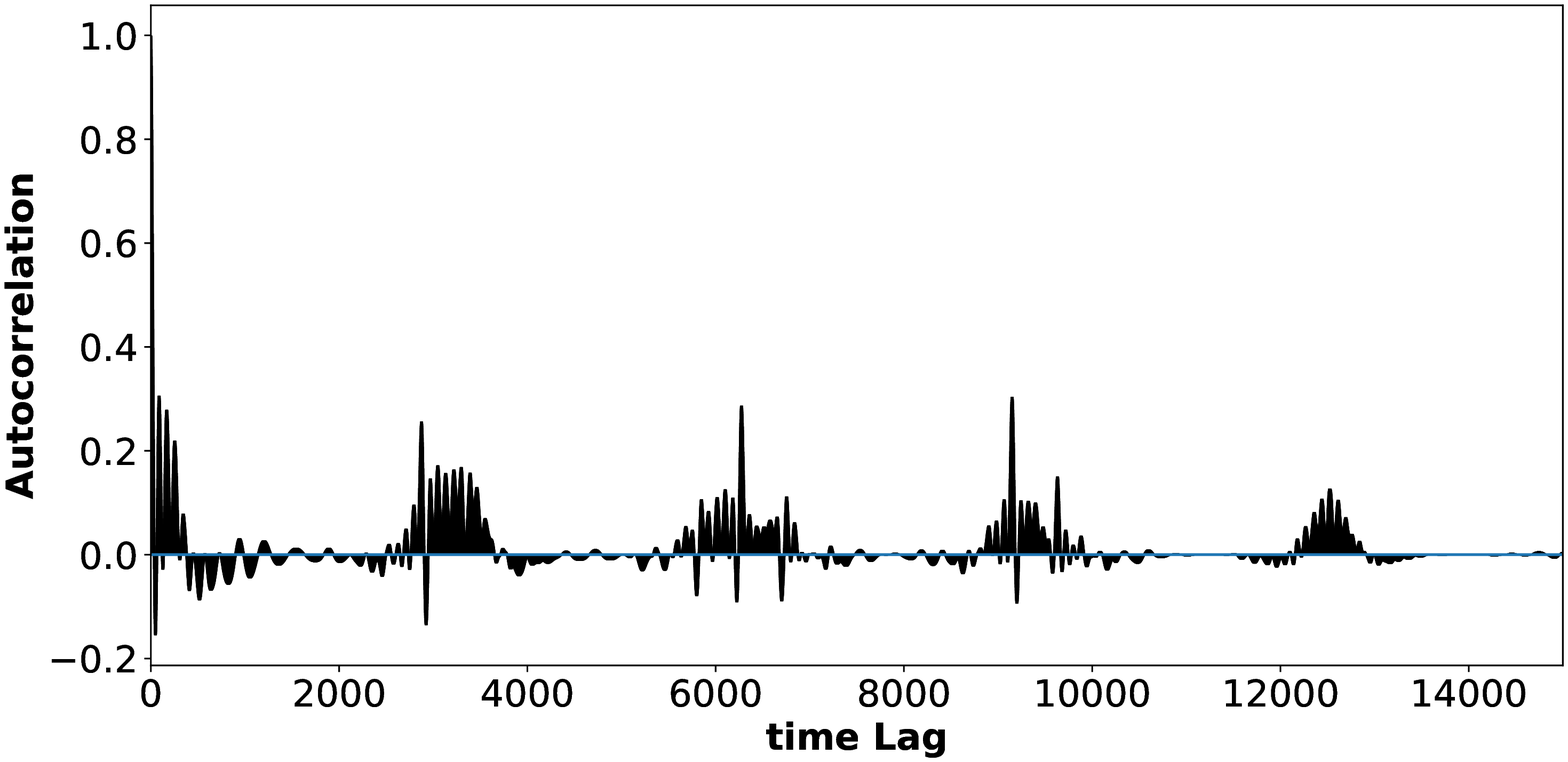}
\caption{Autocorrelation function of $q_{1}(t)$}
\label{toda-xauto_gammaq1}
\end{subfigure}%
\begin{subfigure}{.5\textwidth}
\centering
\includegraphics[width=.8\linewidth]{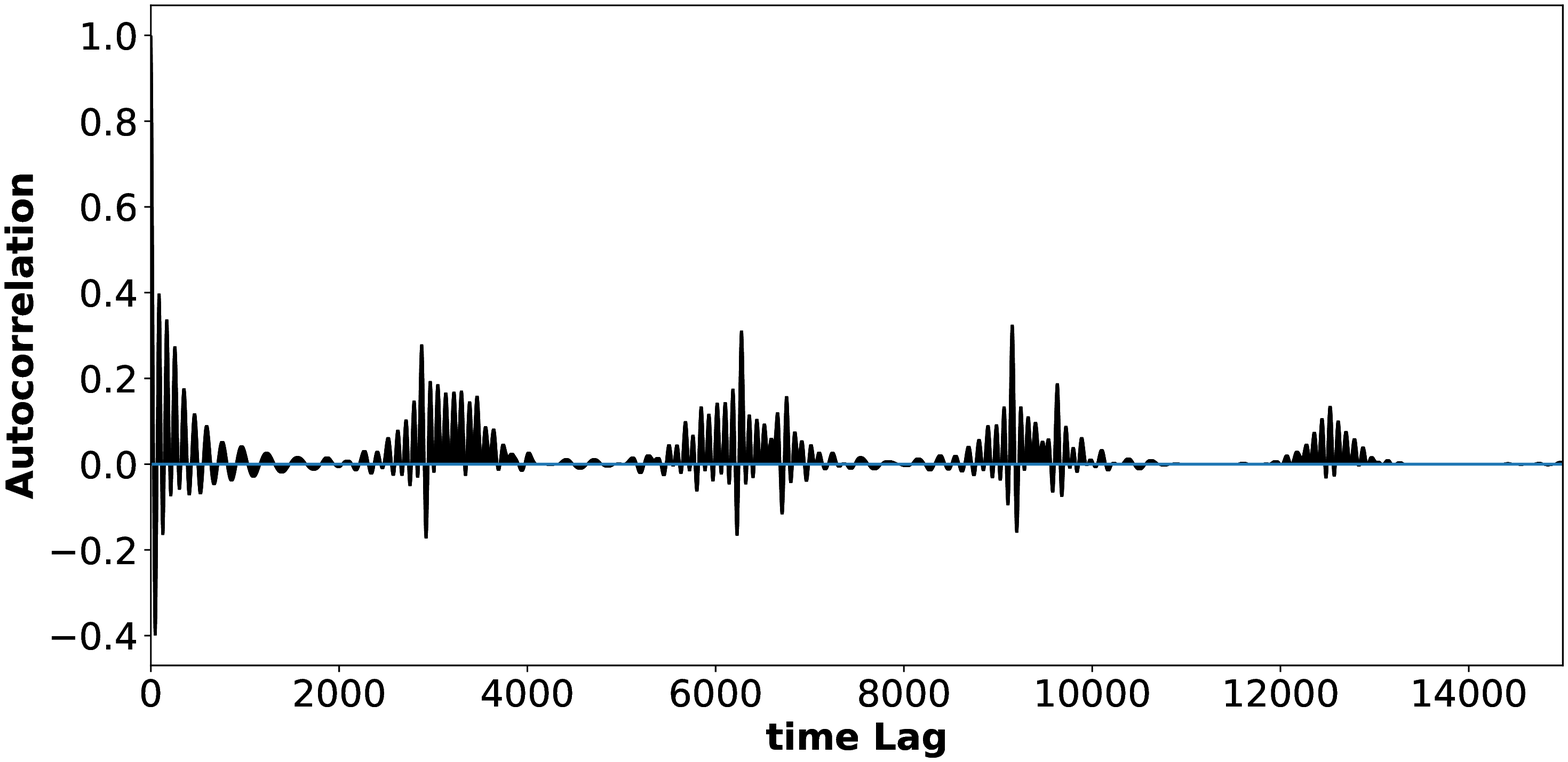}
\caption{Autocorrelation function of $q_{2}(t)$}
\label{toda-xauto_gammaq2}
\end{subfigure}%
\caption{(Color online)  Poincar$\acute{e}$ section, Lyapunov exponents, autocorrelation function
and power spectra for $\Gamma_1=1.85, \Gamma_2=0$ with the initial condition
$Q_{1}(0)=0.001, Q_{2}(0)=0.002, \dot{ Q_{1}}(0)=0.003,\dot{Q_{2}}(0)=0.004$ The plots corresponding to the
largest ($\lambda_1=1.3948$) and the smallest ($\lambda_2=-1.4125$)  Lyapunov exponents
are denoted by blue and violet colors, respectively.} 
\label{multi_toda}
\end{figure}

\begin{figure}
\begin{subfigure}{.5\textwidth}
\centering
\includegraphics[width=.8\linewidth]{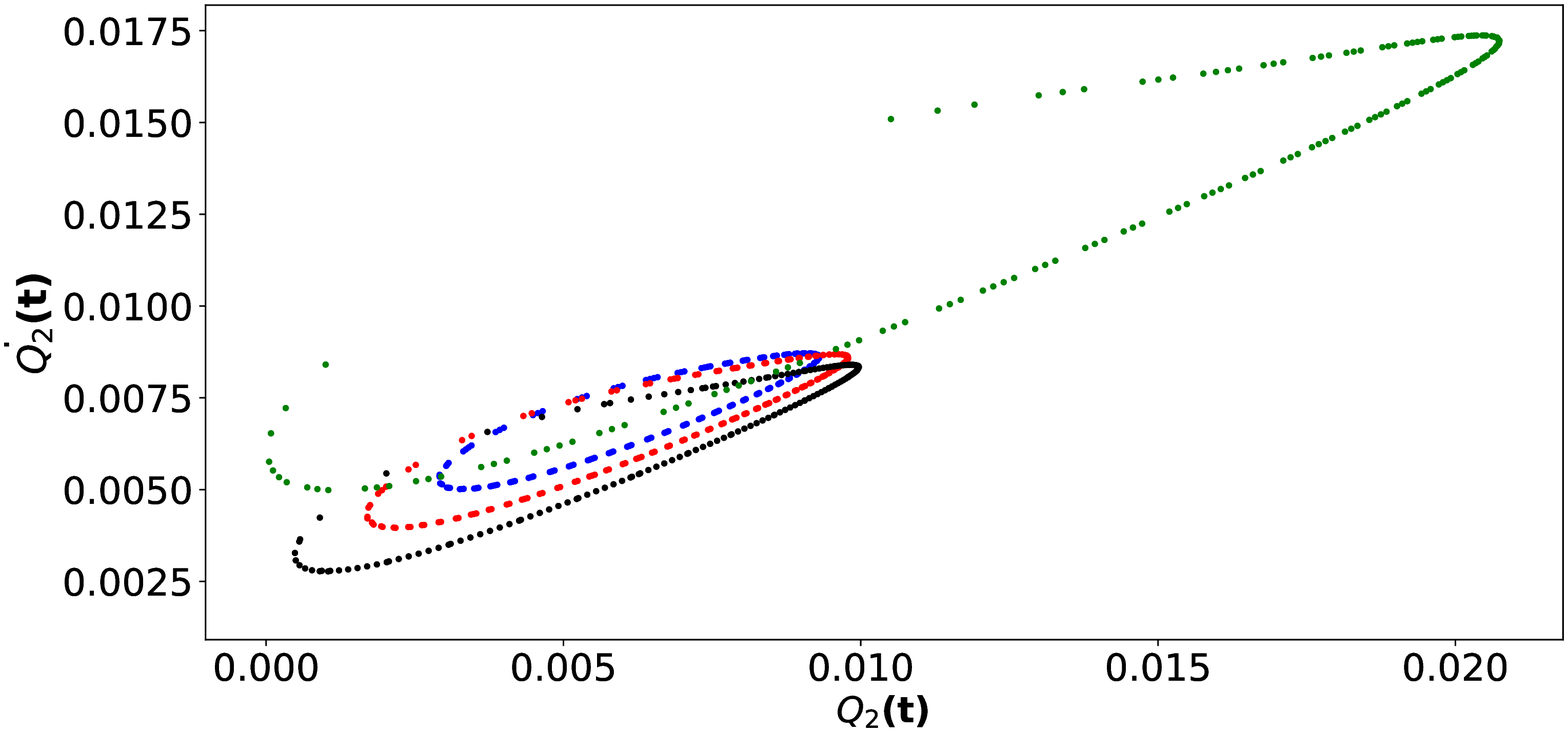} 
\caption{  $\dot{Q_{2}}(t)$ VS. $Q_{2}(t)$ plot($\Gamma_1=2.04, \Gamma_2=0$)}
\label{poincare_periodic}
\end{subfigure}%
\begin{subfigure}{.5\textwidth}
\centering
\includegraphics[width=.8\linewidth]{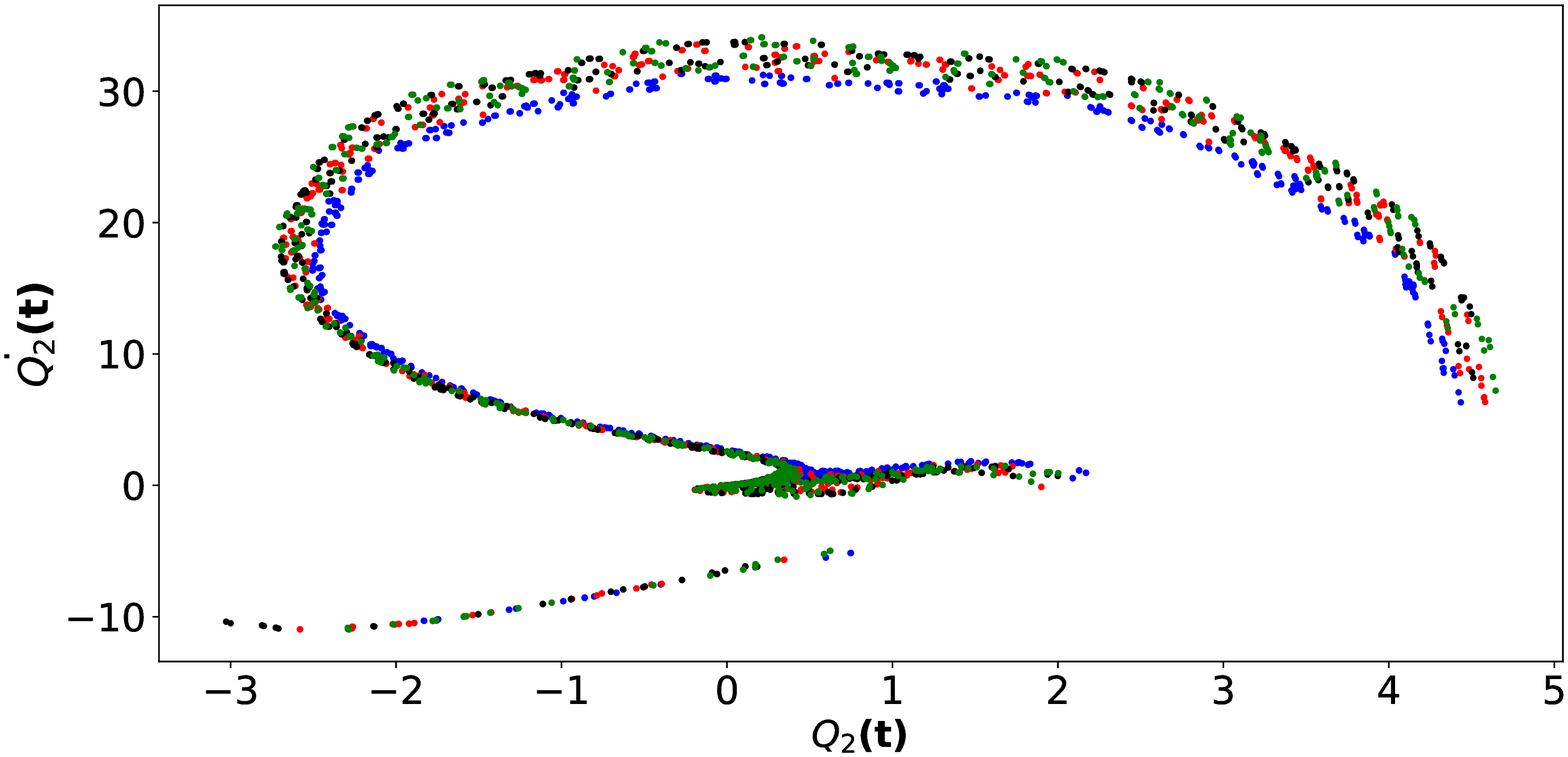}  
\caption{ $\dot{Q_{2}}(t)$ VS. $Q_{2}(t)$ plot( $\Gamma_1=1.90, \Gamma_2=0$)}
\label{poincare_chaotic}
\end{subfigure}%
\caption{(Color online) Poincar$\acute{e}$ section of Eq. ({\ref{eqnn_center_of_mass}}) with four sets of initial
conditions (a) $Q_{1}(0)=0.001,Q_{2}(0)=0.005,\dot{Q_{1}}(0)=0.002,\dot{Q_{2}}(0)=0.004$ (blue color) ,
(b) $Q_{1}(0)=0.001,Q_{2}(0)=0.004,\dot{Q_{1}}(0)=0.002,\dot{Q_{2}}(0)=0.003$  (red color),(c) $Q_{1}(0)=0.001,Q_{2}(0)=0.004,\dot{Q_{1}}(0)=0.003,\dot{Q_{2}}(0)=0.002$  (black color) and (d)$Q_{1}(0)=0.002,Q_{2}(0)=0.005,\dot{Q_{1}}(0)=0.003,\dot{Q_{2}}(0)=0.004$ (green color)
}
\label{poincare_periodic_chaotic}
\end{figure}

We study  Eqn. (\ref{eqnn_center_of_mass}) with $\Gamma_2=\Pi^{(3)}=0$ numerically.
The time-series of the dynamical variables  in the vicinity of the point $Q=0$ is periodic and shown in
Fig. \ref{periodic_ solutions} for $\Gamma_1=2.04$. The Lyapunov exponents and the autocorrelation
functions for the time series representing the periodic solutions in Fig.\ref{periodic_ solutions}
have been calculated to confirm that these solutions are indeed regular.  
The study of sensitivity of the dynamical variables to the initial conditions is one of
the important methods to check whether the system is chaotic or not. The two sets of initial conditions:
(a)$Q_{1}(0)=0.001, Q_{2}(0)=0.002, \dot{ Q_{1}}(0)=0.003,\dot{Q_{2}}(0)=0.004$ and (b)$Q_{1}(0)=0.001,
Q_{2}(0)=0.002, \dot{ Q_{1}}(0)=0.003,\dot{Q_{2}}(0)=0.0045$ are considered in order to study the sensitivity
of the dynamical variables to the initial conditions in different regions of the parameter. It may be noted
that these two initial conditions are identical except for the values of $\dot{Q_2}(0)$ which differ by
$0.005$. The Figs. \ref{Chaotic_solution_Q1_gamma_+ve} and \ref{Chaotic_solution_Q2_gamma_+ve} represent
the time series of the dynamical variables in the chaotic regime for $\Gamma_1=1.85$.
Other independent techniques are also used to support the chaotic behaviour in the system.
In this context, Fig. \ref{multi_toda} shows the auto-correlation function, Lyapunov exponent,
Poincar$\acute{e}$ section, and power spectra for $\Gamma_1 = 1.85$. The Lyapunov exponents are
$(1.3948, 0.71022, -0.68741, -1.4125)$. The largest value of the Lyapunov exponent is observed to
increase with the increasing values of $\Gamma$. The route to chaos has also been studied by investigating
the change in the behaviour of the Poincar\'e sections as a function of $\Gamma_1$. The relevant plots are
given in Fig. \ref{poincare_periodic_chaotic}. The broken orbits in the $Q_2-\dot{Q}_2$ plane in
the chaotic regime are shown in Figs.  \ref{multi-poincare-gamma}  and \ref{poincare_chaotic}, while the
unbroken orbits are given in Fig. \ref{poincare_periodic}.

\begin{figure}[ht!]
\begin{subfigure}{.5\textwidth}
\centering
\includegraphics[width=.8\linewidth]{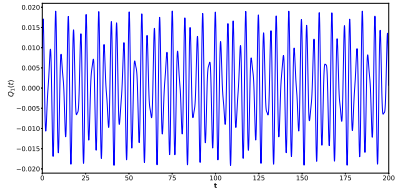} 
\caption{ Regular solution( $\Gamma=1.0$).}
\label{Regular_solution_Q1_h}
\end{subfigure}%
\begin{subfigure}{.5\textwidth}
\centering
\includegraphics[width=.8\linewidth]{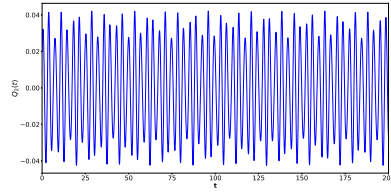} 
\caption{ Regular solution( $\Gamma=1.0$)}
\label{Regular_solution_Q2_h}
\end{subfigure}%
\caption{(Color online)  Regular solutions of Eq. (\ref{eqnn_center_of_mass}) in the vicinity of the point
$P_0$ with the initial conditions $Q_{1}(0)=0.01, Q_{2}(0)=0.02, \dot{ Q_{1}}=0.03,\dot{Q_{2}}=0.04$.}
\label{periodic_solutions-h}
\end{figure}
\begin{figure}[ht!]
\begin{subfigure}{.5\textwidth}
\centering
\includegraphics[width=.8\linewidth]{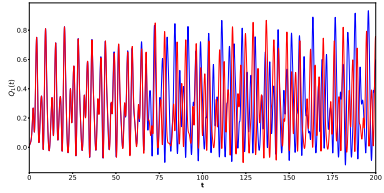} 
\caption{ Chaotic solution( $\Gamma=2.0$).}
\label{Chaotic_solution_Q1_h}
\end{subfigure}%
\begin{subfigure}{.5\textwidth}
\centering
\includegraphics[width=.8\linewidth]{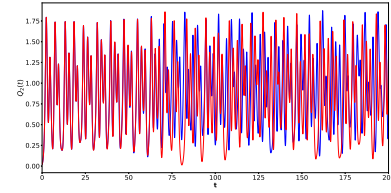} 
\caption{ Chaotic solution( $\Gamma=2.0$)}
\label{Chaotic_solution_Q2_h}
\end{subfigure}%
\caption{(Color online) Chaotic solutions of Toda system with two sets of initial
conditions (a)$Q_{1}(0)=0.01, Q_{2}(0)=0.02, \dot{ Q_{1}}=0.03,\dot{Q_{2}}=0.04$ (blue color) and 
(b)$Q_{1}(0)=0.01, Q_{2}(0)=0.02, \dot{ Q_{1}}=0.03,\dot{Q_{2}}=0.045$(red color).}
\label{chaotic_solutions-h}
\end{figure}

\begin{figure}[ht!]
\begin{subfigure}{.5\textwidth}
\centering
\includegraphics[width=.8\linewidth]{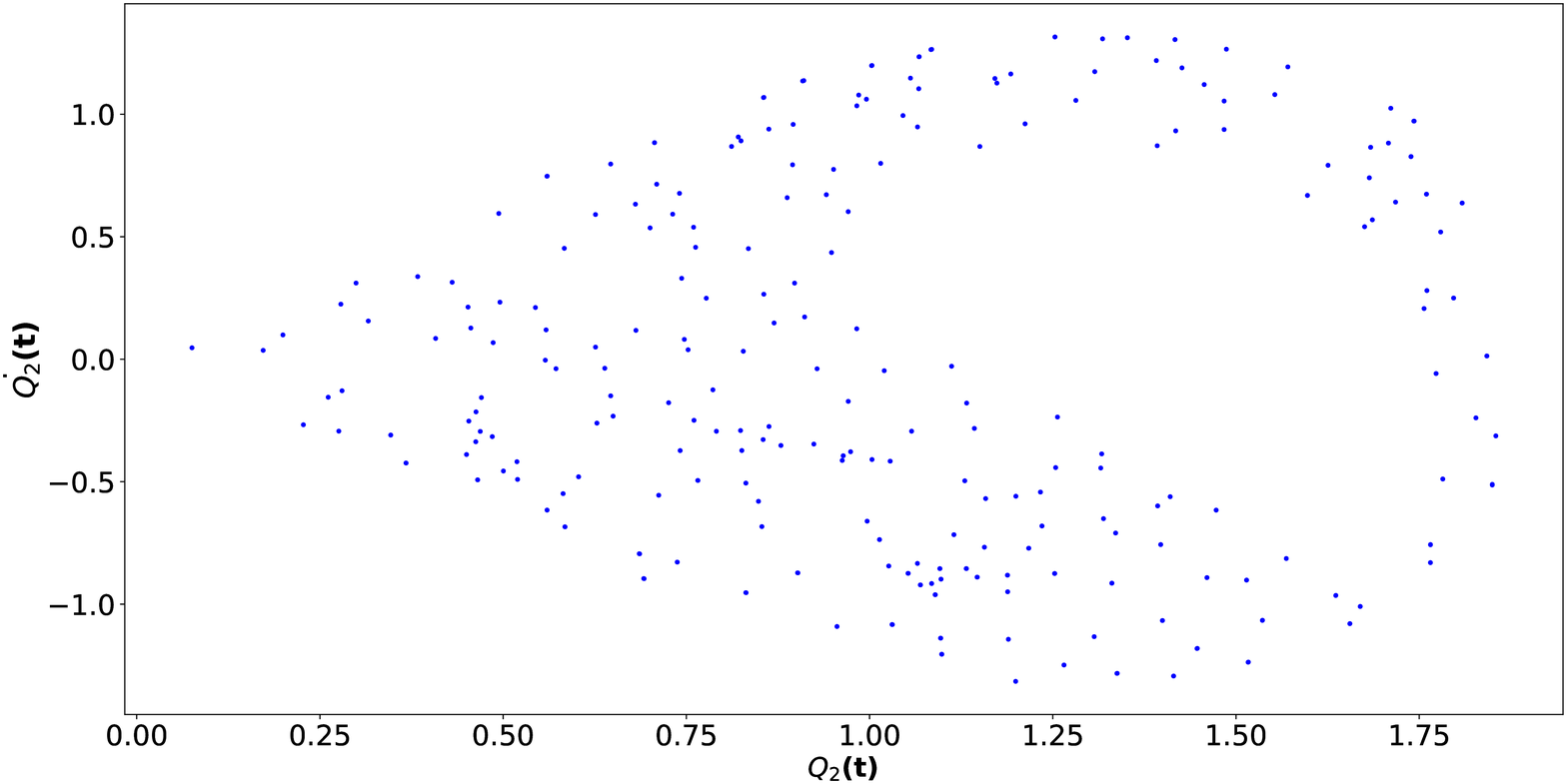}
\caption{Poincar$\acute{e}$ section: $\dot{Q_{2}}(t)$ VS. $Q_{2}(t)$ plot}
\label{multi-poincare-gamma-h}
\end{subfigure}%
\begin{subfigure}{.5\textwidth}
\centering
\includegraphics[width=.8\linewidth]{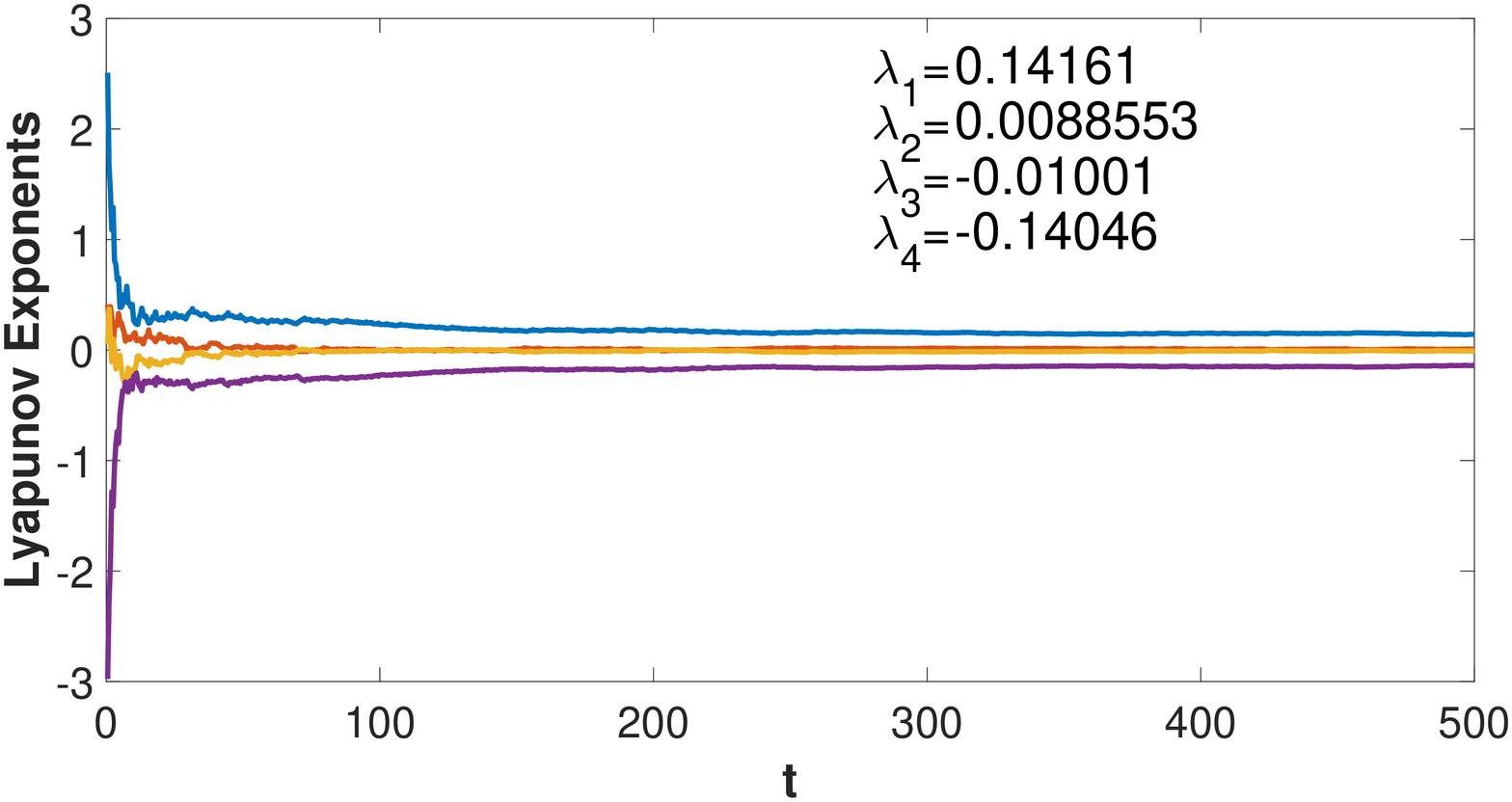}
\caption{Lyapunov exponents}
\label{multi-lyap_h}
\end{subfigure}%
\newline
\begin{subfigure}{.5\textwidth}
\centering
\includegraphics[width=.8\linewidth]{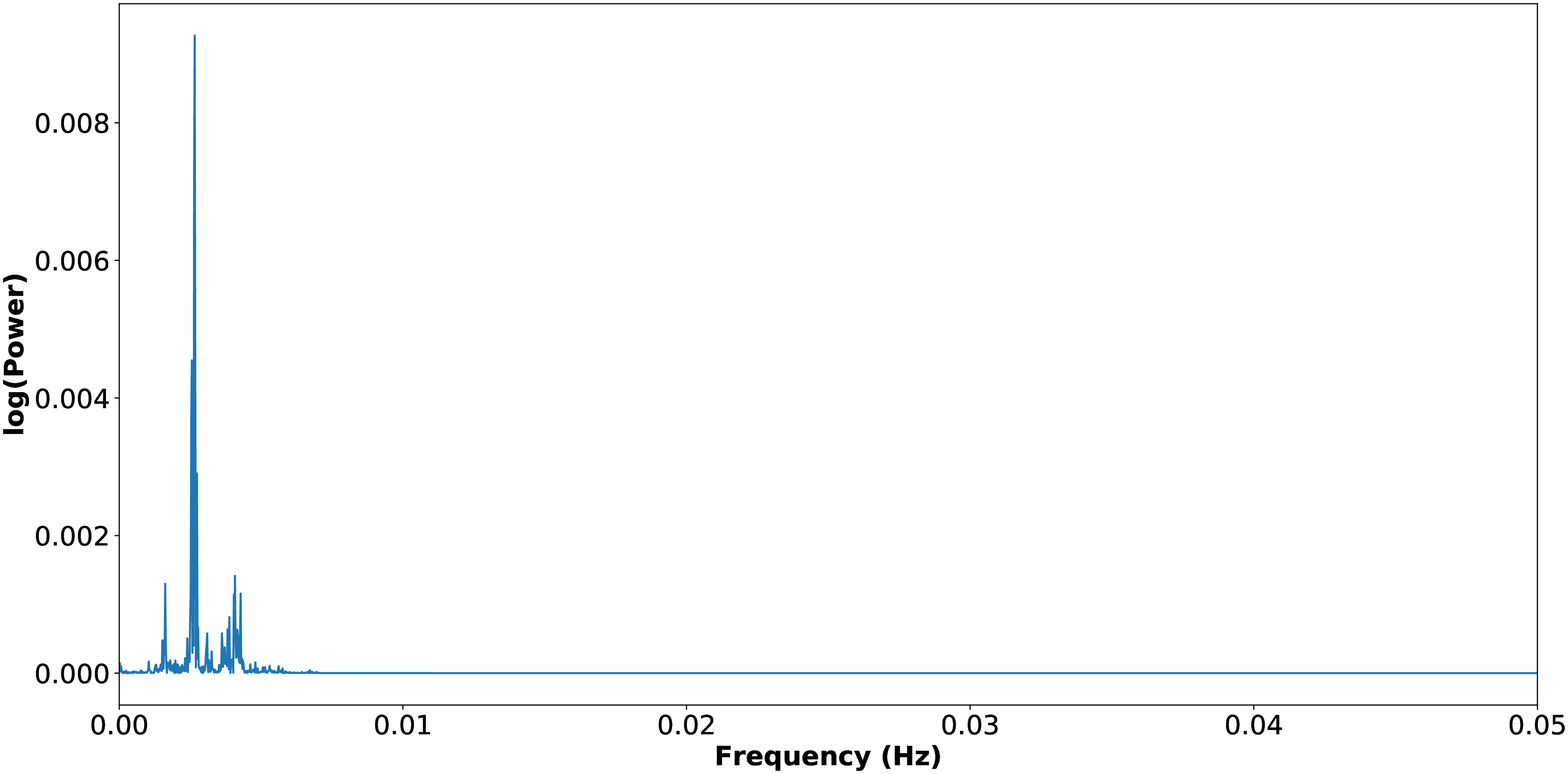}
\caption{Powerspectra  of $q_{1}(t)$}
\label{toda-power_Q1_h}
\end{subfigure}%
\begin{subfigure}{.5\textwidth}
\centering
\includegraphics[width=.8\linewidth]{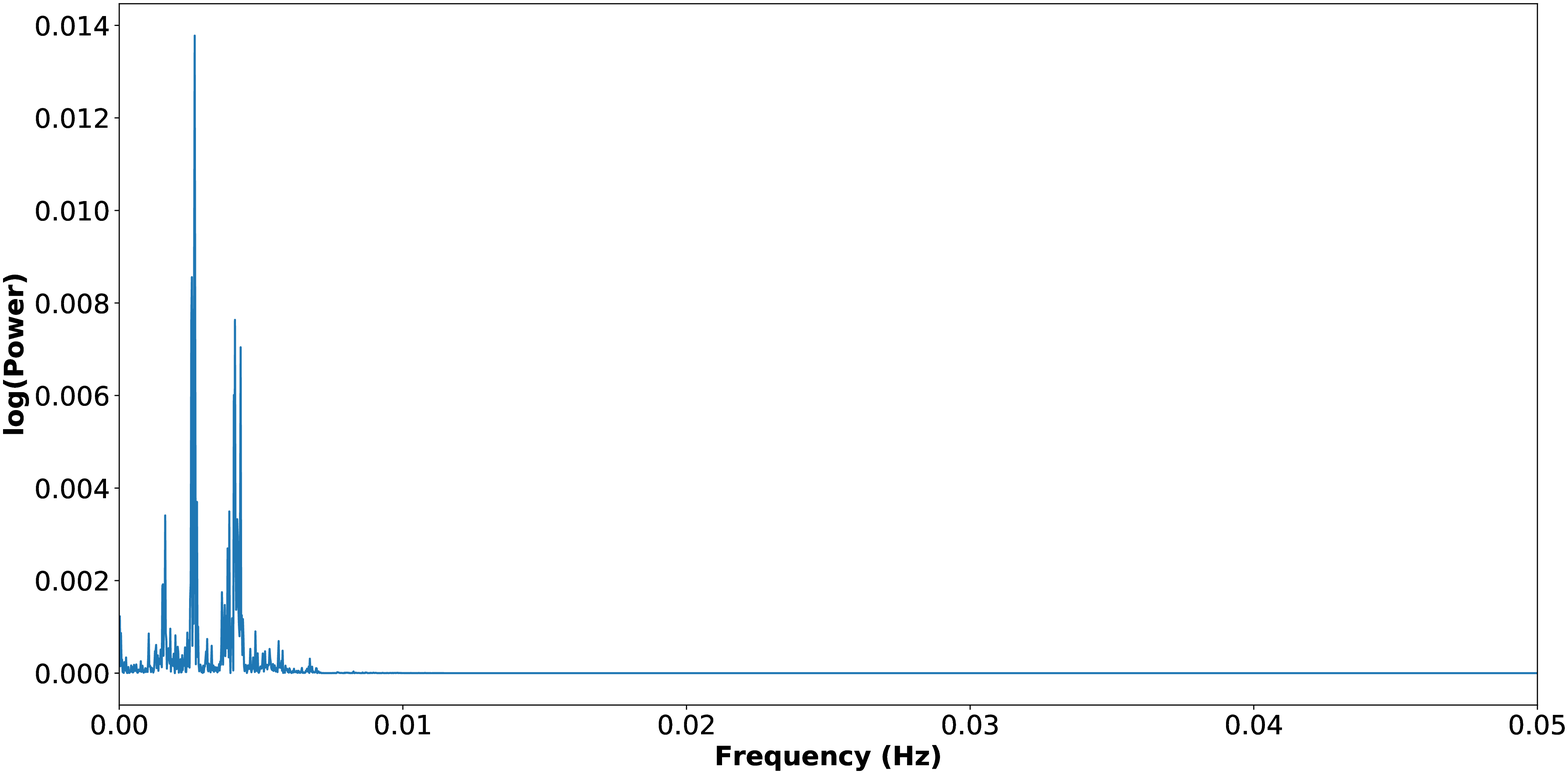}
\caption{Powerspectra  of $q_{2}(t)$}
\label{toda-power_Q2_h}
\end{subfigure}%
\newline
\begin{subfigure}{.5\textwidth}
\centering
\includegraphics[width=.8\linewidth]{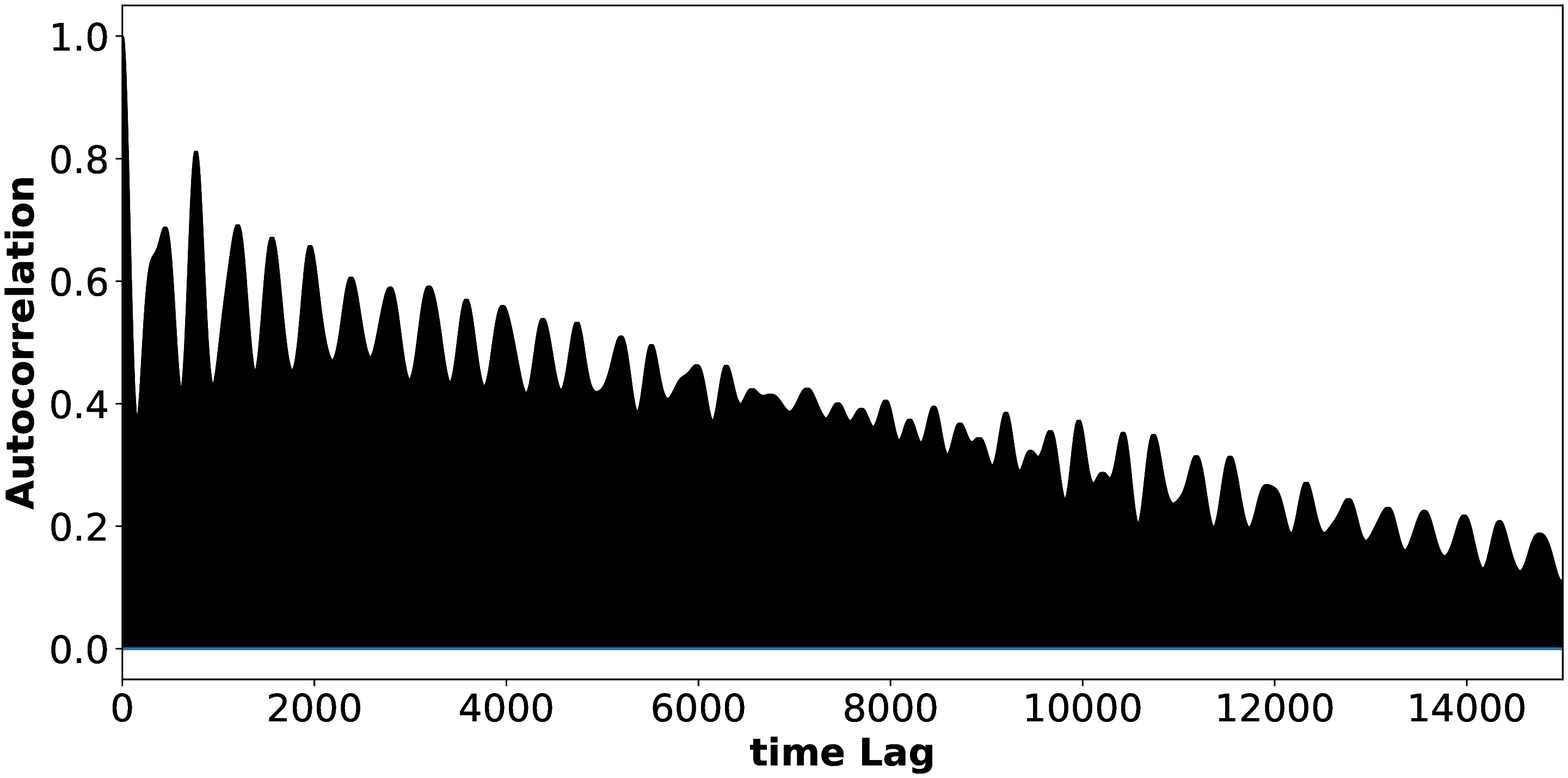}
\caption{Autocorrelation function of $q_{1}(t)$}
\label{toda-xauto_gammaQ1-h}
\end{subfigure}%
\begin{subfigure}{.5\textwidth}
\centering
\includegraphics[width=.8\linewidth]{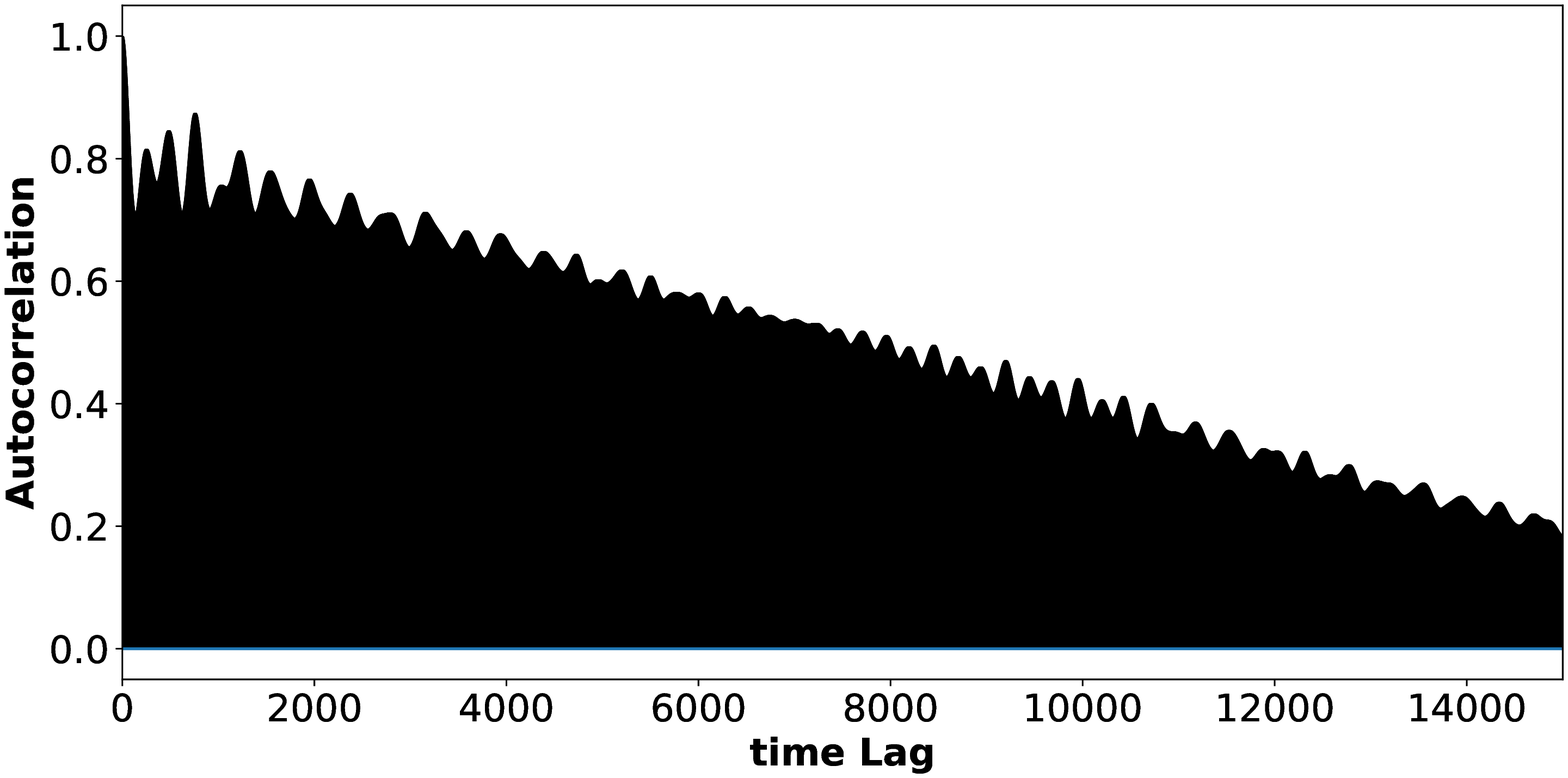}
\caption{Autocorrelation function of $q_{2}(t)$}
\label{toda-xauto_gammaQ2-h}
\end{subfigure}%
\caption{(Color online)  Poincar$\acute{e}$ section, Lyapunov exponents, autocorrelation function
and power spectra for $\Gamma=2.0$ with the initial condition
$Q_{1}(0)=0.01, Q_{2}(0)=0.02, \dot{ Q_{1}}(0)=0.03,\dot{Q_{2}}(0)=0.04$ The plots corresponding to the
largest ($\lambda_1=0.14161$) and the smallest ($\lambda_2=-0.14046$)  Lyapunov exponents
are denoted by blue and violet colors, respectively.} 
\label{multi_toda-h}
\end{figure}

\begin{figure}[ht!]
\begin{subfigure}{.5\textwidth}
\centering
\includegraphics[width=.8\linewidth]{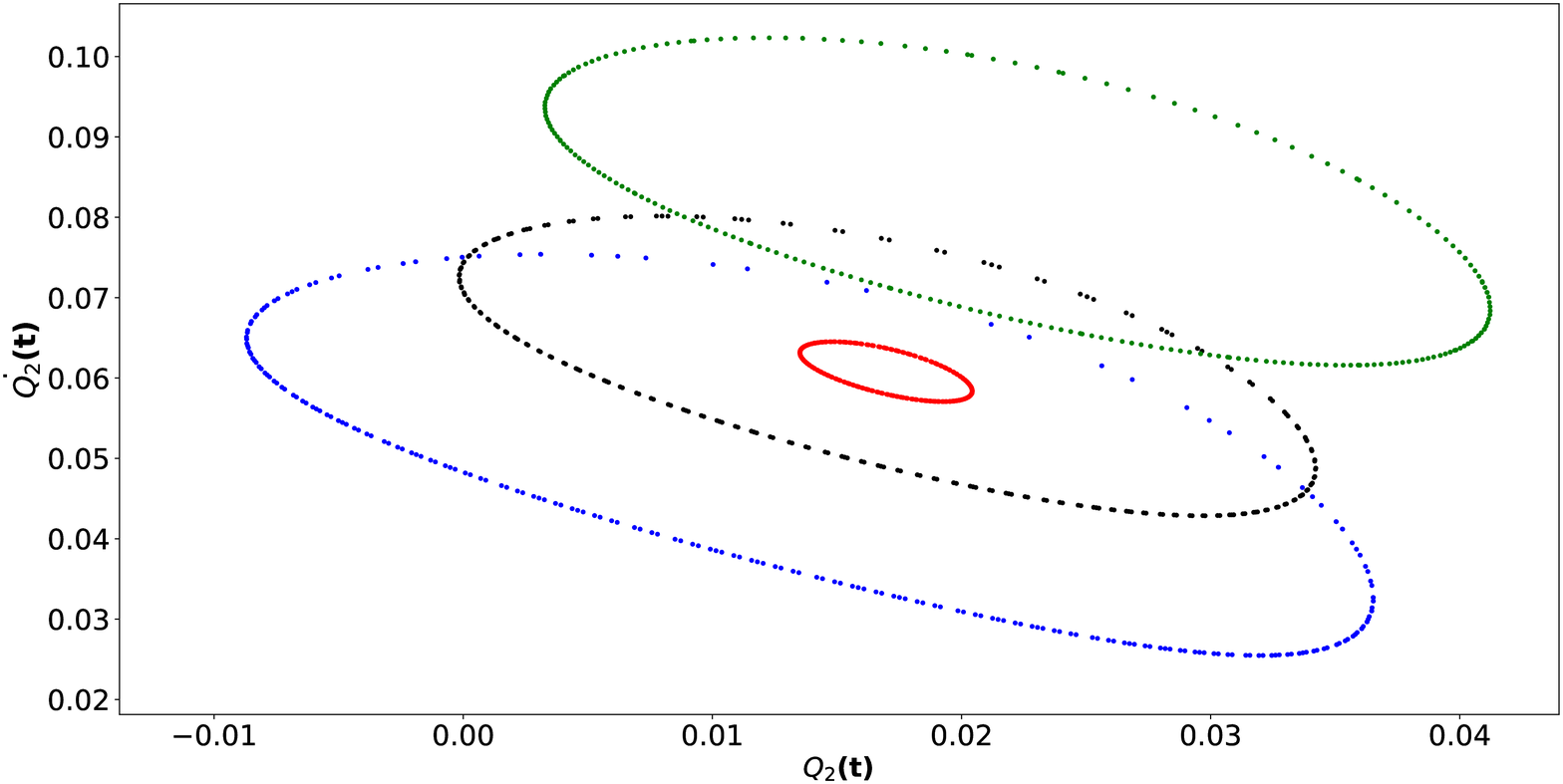} 
\caption{  $\dot{Q_{2}}(t)$ VS. $Q_{2}(t)$ plot($\Gamma=1.2$)}
\label{poincare_periodic-h}
\end{subfigure}%
\begin{subfigure}{.5\textwidth}
\centering
\includegraphics[width=.8\linewidth]{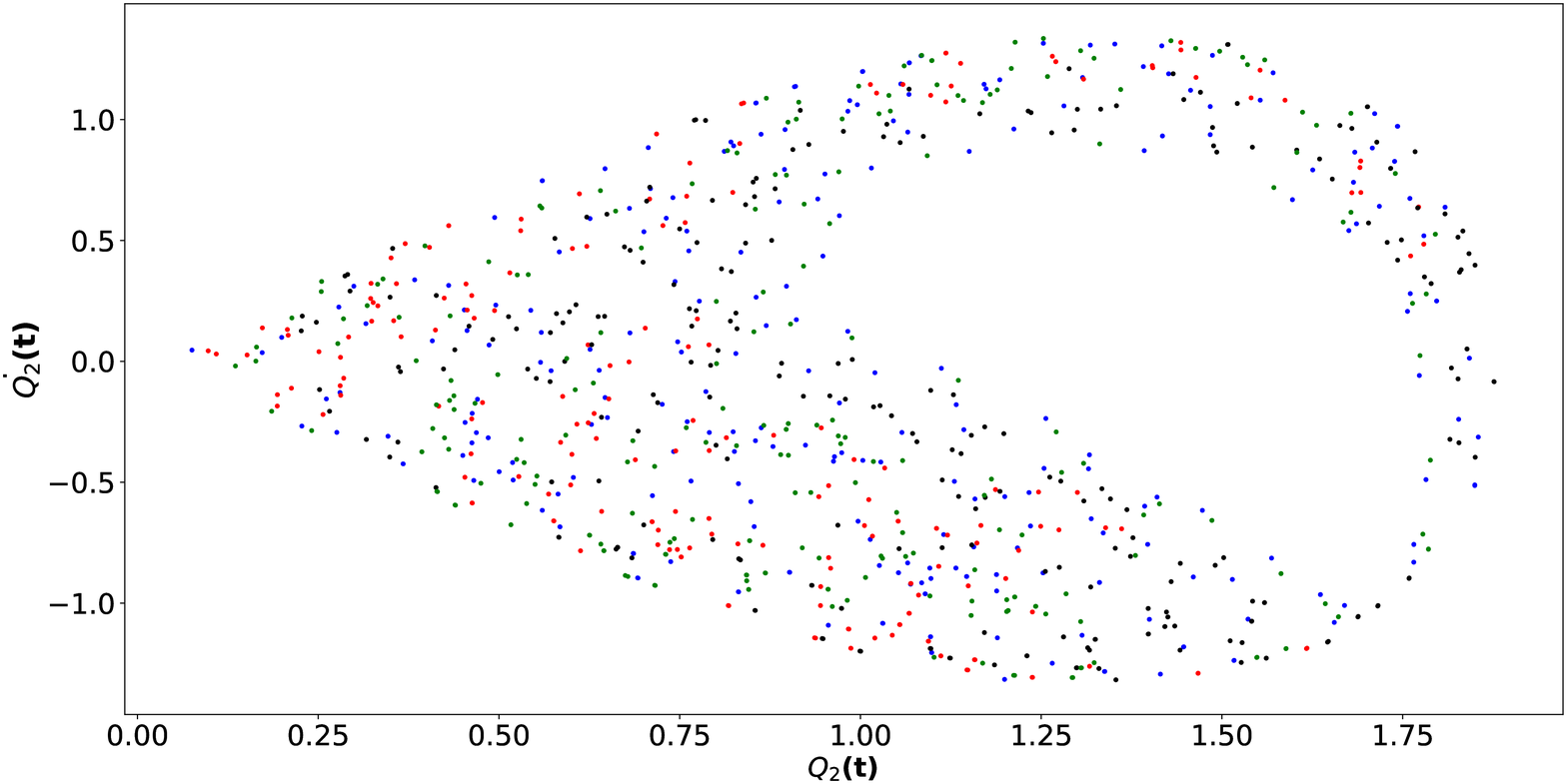}  
\caption{ $\dot{Q_{2}}(t)$ VS. $Q_{2}(t)$ plot( $\Gamma=2.0$)}
\label{poincare_chaotic-h}
\end{subfigure}%
\caption{(Color online) Poincar$\acute{e}$ section of Eq. ({\ref{eqnn_center_of_mass}}) with four sets of initial
conditions (a) $Q_{1}(0)=0.01,Q_{2}(0)=0.05,\dot{Q_{1}}(0)=0.02,\dot{Q_{2}}(0)=0.04$ (blue color) ,
(b) $Q_{1}(0)=0.01,Q_{2}(0)=0.04,\dot{Q_{1}}(0)=0.02,\dot{Q_{2}}(0)=0.03$  (red color),(c) $Q_{1}(0)=0.01,Q_{2}(0)=0.04,\dot{Q_{1}}(0)=0.03,\dot{Q_{2}}(0)=0.02$  (black color) and (d)$Q_{1}(0)=0.02,Q_{2}(0)=0.05,\dot{Q_{1}}(0)=0.03,\dot{Q_{2}}(0)=0.04$ (green color)
}
\label{poincare_periodic_chaotic-h}
\end{figure}

\subsection{Hamiltonian System: Numerical Analysis}

We study  Eqn. (\ref{eqnn_center_of_mass}) numerically for $\Gamma_2=\frac{\Gamma_1}{2}  \neq 0, \Pi^{(3)}=0$. The regular periodic solutions around the equilibrium point $Q=0$ is
shown in Fig. \ref{periodic_solutions-h} for $\Gamma_1=1.0$. The Lyapunov exponents and the autocorrelation
functions for the time series representing the periodic solutions in Fig.\ref{periodic_solutions-h}
have been calculated to confirm that these solutions are indeed regular.  
The chaotic behaviour in the system has been investigated by studying sensitivity of the dynamical
variables to the initial conditions. The two sets of initial conditions:
(a)$Q_{1}(0)=0.01, Q_{2}(0)=0.02, \dot{ Q_{1}}(0)=0.03,\dot{Q_{2}}(0)=0.04$ and (b)$Q_{1}(0)=0.01,
Q_{2}(0)=0.02, \dot{ Q_{1}}(0)=0.03,\dot{Q_{2}}(0)=0.045$ are considered in order to study the sensitivity
of the dynamical variables to the initial conditions in different regions of the parameter. It may be noted
that these two initial conditions are identical except for the values of $\dot{Q_2}(0)$ which differ by
$0.05$. The Fig. \ref{chaotic_solutions-h} represents the time series of the dynamical variables in
the chaotic regime for $\Gamma_1=2.0$.
Other independent techniques are also used to support the chaotic behaviour in the system.
In this context, Fig. \ref{multi_toda-h} shows the auto-correlation function, Lyapunov exponent,
Poincar$\acute{e}$ section, and power spectra for $\Gamma_1 = 2.0$. The Lyapunov exponents are
$(0.14161, 0.0088553, -0.01001, -0.14046)$. The largest value of the Lyapunov exponent is observed to
increase with the increasing values of $\Gamma_1$.  The route to chaos is  studied by investigating
the change in the behaviour of the Poincar\'e sections as a function of $\Gamma_1$. The transition from
broken to unbroken orbits in the $Q_2-\dot{Q}_2$-plane is seen as $\Gamma_1$ is increased beyond $\Gamma_1=2$.
The broken orbits in the $Q_2-\dot{Q}_2$ plane in
the chaotic regime are given in Figs.  \ref{multi-poincare-gamma-h}  and \ref{poincare_chaotic-h}, while the
unbroken orbits are shown in Fig. \ref{poincare_periodic-h}.

\section{Conclusions $\&$ Discussions}

We have studied periodic Toda oscillators with balanced loss-gain for two and three particles.
The two-particle system is integrable. The system is invariant under both space and time
translation. The integrals of motion are the Hamiltonian and a  generalized total momentum.
In the center of mass coordinate, the system reduces to that of a particle moving in a
harmonic oscillator plus cosine hyperbolic potential. The nature of the effective
potential is that of a single-well for $\Gamma < \frac{1}{2}$ and a double-well for
$\Gamma > \frac{1}{2}$. The double-well potential is symmetric or asymmetric depending
on whether the generalized total momentum is vanishing or non-vanishing, respectively.
Exact analytic solutions may be found in some limiting cases and periodic solutions exist within some ranges of the
loss-gain strength. The system in its full generality has been investigated numerically and
periodic solutions have been presented.

The three-particle Toda oscillators with balanced loss-gain is non-Hamiltonian and regular
periodic solutions exist only for a narrow range of the strength of the balanced loss-gain.
The system can be made Hamiltonian by introducing additional velocity mediated coupling
among the particles with its strength being half of the strength of the loss-gain.
The region of the regular periodic solution enhances significantly compared to the
case with vanishing velocity mediated coupling. We have presented the linear stability
analysis and bifurcation diagram for a system with arbitrary strengths for the balance
loss-gain and velocity mediated coupling. The system is chaotic in most of the regions
in the space of two parameters characterizing the strengths of the loss-gain and
velocity-mediated coupling. Although the generic system is non-Hamiltonian,
the generalized total momentum is an integral of motion which is a consequence of the
translation invariance. The regular and chaotic dynamics of the system has been studied
for the Hamiltonian system and non-Hamiltonian system with velocity mediated
coupling. The chaotic behaviour has been studied in detail through sensitivity of
the solution to the initial conditions, Poincar${\acute{e}}$-sections, auto-correlation
functions and power-spectra. The Lyapunov exponents have been  computed numerically
and the largest Lyapunov exponent is seen to increase with increasing values of the
loss-gain strength. The quasi-periodic route to chaos has also been investigated numerically.

The equal-mass Toda lattice without any driving term is non-chaotic. We have shown that
the inclusion of balanced loss-gain for equal-mass Toda oscillators without any driving
term is chaotic. Further, we have presented Hamiltonian chaos within the context of
Toda lattice with balanced loss-gain and velocity mediated coupling. It is expected that further investigations on this
new class of systems from various angles will provide new information on dynamical
systems that has not yet been explored. For example, an inhomogeneous Toda lattice
with balanced loss-gain mimicking realistic physical scenario may be investigated\cite{pp}. Further,
a few particular directions of physical interests may be to investigate the quantized
versions of these systems along with studies on quantum chaos, quantum criticality,
quantum synchronizations etc. Some of these issues will be addressed in future.

\section{Acknowledgements}

The work of PR is partly supported by CSIR-NET fellowship({\bf CSIR File No.:  09/202(0072)/2017-EMR-I})
of Govt. of India.


\begin{thebibliography}{10}

\bibitem{mt} M.Toda, Studies of a non-linear lattice, Phys. Rep. {\bf 18}, 1 (1975).

\bibitem{toda-book}  M. Toda, Theory of Nonlinear Lattices, 2nd enl. ed., Springer, Berlin, 1989.

\bibitem{flascha} H. Flaschka, The Toda lattice. II. Existence of integrals, Phys.
	Rev. {\bf B 9}, 1924 (1974).

\bibitem{mh} M. H\'enon, Integrals of the Toda lattice, Phys. Rev. {\bf B9}, 1921(1974)

\bibitem{mpc} M. A. Agrotis, P.A.Damianou and  C.Sophocleous, The Toda lattice is super-integrable,
Physica A {\bf 365}, 235(2006).

\bibitem{kac} M. Kac and P. Van Moerbeke, A complete solution of the periodic Toda problem,
Proc. Natl. Acad. Sci. USA {\bf 72}, 2879 (1975).

\bibitem{tanaka} E. Date and S. Tanaka, Periodic multi-soliton solutions of
Korteweg-de Vries equation and Toda lattice, Prog. Theor. Phys.
{\bf 59}, 107 (1976).

\bibitem{ac} A. Cuccoli et al., Thermodynamics of the Toda Chain, Int. J.
	Mod. Phys. {\bf B 08}, 2391 (1994).

\bibitem{heat} M. Toda, Solitons and Heat Conduction, Phys. Scr. {\bf 20}, 424(1979).

\bibitem{heat-1} M. Sataric, J. A. Tuszynski, R. Zakula, and S. Zekovic,
Heat conductivity of a perturbed monatomic Toda lattice without impurities,
J. Phys.: Condens. Matter {\bf 6}, 3917 (1994).

\bibitem{dna} V. Muto, A. C. Scott, and P. L. Christiansen, A Toda lattice model for DNA:
Thermally generated solitons, Physica D {\bf 44}, 75 (1990).

\bibitem{helix} F. d'Ovivio, H. G. Bohr, and P. Lindgard, Solitons on H Bond in Proteins,
J. Phys.: Condens. Matter {\bf 15}, S1699 (2003).


\bibitem{ga}  G.L.Oppo, and A.Politi, Toda potential in laser equations. Z. Phys. B - Condensed Matter 59, 111–115 (1985);
\bibitem{ysmj} Y. Lien, S. M. de Vries, M. P. van Exter, and J. P. Woerdman, Lasers as Toda oscillators,
J. Optical Soc. Am. B {\bf 19(6)}, 1461–1466 (2002).
\bibitem{scfc} Simone Cialdi, Fabrizio Castelli and Franco Prati,
Lasers as Toda oscillators: An experimental confirmation, Optics Communications {\bf 287}, 176 (2013).


\bibitem{casati} G. Casati and J. Ford, Stochastic transition in the unequal-mass Toda lattice,
	Phys. Rev. A {\bf 12}, 1702(1975).
\bibitem{inhomo} L. Vergara and B. A. Malomed, Suppression of the generation of defect
modes by a moving soliton in an inhomogeneous Toda lattice, Phys. Rev.  E {\bf 77},
047601 (2008).

\bibitem{habib} S. Habib, H. E. Kandrup and M. E. Mahon, Chaos and noise in a
truncated Toda potential, Phys. Rev. E {\bf 53}, 5473(1996).

\bibitem{topo} M. Ezawa, Topological Edge States and Bulk-edge Correspondence in
Dimerized Toda Lattice, J. Phys. Soc. Jpn. 91, 024703 (2022).

\bibitem{dissi1} J Hietarinta, T Kuusela and B A Malomed, Shock waves in the
dissipative Toda lattice, J. Phys. A: Math. Gen. {\bf 28}, 3015(1995).

\bibitem{dissi2} K. \O. Rasmussen, Boris A. Malomed, A. R. Bishop, and Niels Gr\o nbech-Jensen,
Soliton motion in a parametrically ac-driven damped Toda lattice, Phys. Rev. E {\bf 58},
6695(1998).

\bibitem{dissi3} W. Ebeling, U. Erdmann, J. Dunkel  and M. Jenssen,
Nonlinear Dynamics and Fluctuations of Dissipative Toda Chains, Journal of Statistical Physics,
{\bf 101}, 443 (2000). 

\bibitem{dissi4} V. A. Makarov, E. del Rio, W. Ebeling and M. G. Velarde,
Dissipative Toda-Rayleigh lattice and its oscillatory modes, Phys. Rev. E {\bf 64},
036601 (2001).

\bibitem{dissi5} Alexis Arnaudon, Structure preserving noise and dissipation in the Toda lattice,
J. Phys. A: Math. Theor. {\bf 51}, 214001(2018).








\bibitem{pkg-review} P. K. Ghosh, Classical Hamiltonian Systems with balanced loss
and gain, J. Phys.: Conf. Ser. {\bf 2038}, 012012(2021)

\bibitem{bpeng}B. Peng, S. K. Ozdemir, F. Lei, F. Monifi, M. Gianfreda,
G. L. Long, S. Fan, F. Nori, C. M. Bender, and L. Yang,
Parity-time-symmetric whispering-gallery microcavities,
Nature Physics {\bf10}, 394 (2014).

\bibitem{ben} C. M. Bender, M. Gianfreda, S. K. Ozdemir, B. Peng, and
L. Yang, Twofold transition in PT-symmetric coupled oscillators,
Phys. Rev. A {\bf88}, 062111 (2013).


\bibitem{ds-pkg} D. Sinha, P. K. Ghosh,
${\cal{PT}}$-symmetric rational Calogero model with balanced loss and gain,
Eur. Phys. J. Plus {\bf 132}, 460 (2017).

\bibitem{ben1} C. M. Bender, M. Gianfreda and S. P. Klevansky,
Systems of coupled PT-symmetric oscillators,
Phys. Rev A {\bf90}, 022114 (2014).

\bibitem{khare-0} Jesús Cuevas, Panayotis G. Kevrekidis, Avadh Saxena and
Avinash Khare, PT-symmetric dimer of coupled nonlinear oscillators,
Phys. Rev. A {\bf 88}, 032108 (2013). 

\bibitem{ivb} I. V. Barashenkov and M. Gianfreda,
An exactly solvable $\mathcal {PT}$-symmetric dimer from a Hamiltonian
system of nonlinear oscillators with gain and loss,
J.  Phys. A: Math. Theor. 47, 282001(2014).


\bibitem{khare} A. Khare and A. Saxena,
Integrable oscillator type and Schrödinger type dimers,
J. Phys. A: Math. Theor. {\bf50}, 055202 (2017).

\bibitem{pkg-ds} P. K. Ghosh and Debdeep Sinha, 
Hamiltonian formulation of systems with balanced loss-gain and exactly
solvable models,
Annals of Physics {\bf 388}, 276 (2018).

\bibitem{ds-pkg1} D. Sinha, P. K. Ghosh, On the bound states and correlation
functions of a class of Calogero-type quantum many-body problems with balanced
loss and gain, J. Phys. A: Math. Theor. {\bf 52}, 505203 (2019). 

\bibitem{p6-deb}  D. Sinha and P. K. Ghosh, Integrable coupled 
Li$\acute{e}$nard-type systems with balanced loss and gain, 
Annals of Physics {\bf 400}, 109 (2019).

\bibitem{pkg-pr} Pijush K. Ghosh and Puspendu Roy, On regular and chaotic dynamics of a non-${\cal{PT}}$-symmetric 
Hamiltonian system of a coupled Duffing oscillator with balanced loss and gain, J. Phys.  A: Math. Theor. {\bf 53},
(2020) 475202.
\bibitem{pr-pkg} Puspendu Roy and Pijush K. Ghosh, Complex dynamical properties of coupled Van der Pol-Duffing oscillators with balanced loss and gain, J. Phys. A: Math. Theor. {\bf 55}, 315701(2022).

\bibitem{pkg-1} Pijush K. Ghosh, Taming Hamiltonian systems with balanced loss
and gain via Lorentz interaction : General results and a case study with Landau
Hamiltonian, J. Phys.  A: Math. Theor. {\bf 52}, 415202(2019).

\bibitem{kono} Vladimir V. Konotop, Jianke Yang, and Dmitry A. Zezyulin, Nonlinear waves in PT-symmetric systems,
	Rev. Mod. Phys. {\bf 88}, 035002
\bibitem{sg} Supriyo Ghosh and Pijush K. Ghosh,
Non-linear Schr\"odinger equation with time-dependent balanced loss-gain and space-time modulated non-linear interaction,
e-Print: 2112.04802 [math-ph]; Supriyo Ghosh and Pijush K. Ghosh, Solvable Limits of a class of generalized Vector Nonlocal
Nonlinear Schr\"odinger equation with balanced loss-gain, e-Print: 2212.02786 [nlin.SI].
\bibitem{vsa} V. S. Afraimovich, N. N. Verichev, and M. I. Rabinovich, Stochastic synchronization of
oscillation in dissipative system, Radiophysics and Quantum Electronics {\bf 29}, 795(1986).
\bibitem{ivan} M.V. Ivanchenko, G.V. Osipov, V.D. Shalfeev and J. Kurths, Synchronization of two non-
scalar-coupled limit-cycle oscillators, Physica D {\bf 189}, 8(2004).
\bibitem{apk} A. P. Kuznetsov and J. P. Roman, Properties of synchronization in the systems of
non-identical coupled van der Pol and van der Pol-Duffing oscillators:Broadband synchronization.
Physica D {\bf 238}, 1499(2009).
\bibitem{bpm} B. P. Pandey and Mark Wardle, Hall magnetohydrodynamics of partially ionized plasmas,
Mon. Not. R. Astron. Soc. {\bf 385}, 22692278 (2008).
\bibitem{pp} Puspendu Roy and Pijush K. Ghosh, Inhomogeneous Toda lattice with balanced loss-gain, under preparation.
\end{thebibliography}
\end{document}